\documentclass[aps]{revtex4-1}

\usepackage{bm}
\usepackage{graphicx}
\usepackage{amsmath}
\usepackage{amssymb}

\def\d{\partial}

\begin{document}

\title{Storage of phase-coded patterns via STDP in
fully-connected and sparse network: a study of the network capacity.}

\author{S. Scarpetta$^{1,3}$, A. de Candia$^{2,3,4}$, F. Giacco$^{1,3}$}

\affiliation{%
$^1$Dipartimento di Fisica ``E.R.Caianiello'', Universit\`a di Salerno, Fisciano (SA), Italy
$^2$Dipartimento di Scienze Fisiche, Universit\`a di Napoli Federico II\\
$^3$INFN, Sezione di Napoli e Gruppo Coll. di Salerno\\
$^4$CNR-SPIN, Unit\`a di Napoli\\
}

\begin{abstract}
We study the storage and retrieval of phase-coded patterns as
stable dynamical attractors in recurrent neural networks,
for both an analog and a integrate-and-fire spiking model.
The synaptic strength is determined by a learning rule
based on spike-time-dependent plasticity, with an asymmetric
time window depending on the relative timing between pre- and post-synaptic
activity.  We store multiple patterns and study the network capacity.

For the analog model, we find that the network capacity scales linearly with the network size,
and that both capacity and the oscillation frequency
of the retrieval state depend on the asymmetry of the learning time window.
In addition to fully-connected networks, we study sparse networks,
where each neuron is connected only to a small number $z\ll N$ of other neurons.
Connections can be short range, between neighboring neurons
placed on a regular lattice, or long range, between randomly chosen
pairs of neurons.
We find that a small fraction of long range connections is able to amplify
the capacity of the network. This imply that a
small-world-network topology
is  optimal, as a compromise between the cost of long range connections and the
capacity increase.

Also in the spiking integrate and fire model the crucial result of storing and retrieval of
multiple phase-coded patterns is observed. The capacity of the fully-connected spiking network is
investigated, together with the relation between oscillation frequency of retrieval state and window asymmetry.
\end{abstract}

\maketitle

Recent advances in  brain research have generated renewed awareness and
appreciation that the brain operates as a complex nonlinear dynamic system,
and synchronous and phase-locked oscillations may play a crucial role in information
processing, such as  feature grouping, saliency  enhancing
\cite{Singer99, Fries2005, Fries2007}
and phase-dependent coding of objects in short term memory \cite{MillerPNAS}.
Many results led to the conjecture that synchronized and
phase locked oscillatory neural activity play a fundamental role in
perception, memory, and sensory computation \cite{science,burgess,Gelperin}.

There is increasing evidence
that information encoding  may depend on the temporal
dynamics between neurons, namely, the specific phase alignment
of spikes relative to rhythmic activity across the neuronal
population (as reflected in the local field potential, or LFP)
 \cite{MillerPNAS,8,9,10,11,12}.
Indeed phase-dependent  coding, 
 that exploits the precise temporal relations between the discharges of neurons, may be an effective 
strategy to encode information \cite{Singer99,MillerPNAS,12,13,NC,PREYoshioka}.
 Data from rodents  %
indicate that spatial information may be encoded at specific
phases of ongoing population theta oscillations in the hippocampus\cite{8}, and that spike sequences 
can be replayed at a different time scale \cite{replay3},
while data from monkeys \cite{MillerPNAS} show that phase coding may be a more general coding scheme.

The existence of a periodic spatio-temporal pattern of precisely timed spikes, as attractor of neural dynamics,
has been investigated in different recurrent neural models \cite{SRM,Borisyuk,Jin2002,Timme2006a,Timme2006b,14}.
In Ref.\ \cite{Jin2002} it was shown that periodic spike sequences are attractors of the dynamics,
and the stronger the global inhibition of the network, the faster the rate of convergence to a periodic pattern.
In Refs.\ \cite{Timme2006a,Timme2006b} the problem of finding the set of all networks that exhibit a predefinite
periodic precisely timed pattern was studied.

Here we study how a STDP learning rule can encode many different periodic patterns in a recurrent network,
in such a way that
a pattern can be retrieved initializing the network in a state similar to it, or inducing a short train
of spikes extracted from the pattern.

In our model, information about an item is encoded in the specific phases of firing,
and each item corresponds to a different pattern of phases among units.
Multiple items can be memorized in the synaptic connections, and the 
intrinsic network dynamics recall the specific phases of firing when a partial cue is presented.

Each item with specific phases of firing corresponds to specific relative timings between neurons.
Therefore it seems that phase coding may be well suited to facilitate long-term
storage of items by means of spike-timing dependent plasticity (STDP) \cite{markram,biandpoo,biandpoo2}.

Indeed experimental findings on STDP 
further underlined the importance of precise
temporal relationships of the dynamics,
by showing  that long term changes
in synaptic strengths depend on the precise relative timing of pre- and
post-synaptic spikes  \cite{magee, debanne, biandpoo, biandpoo2, markram,  feldman}.

The computational role and functional implications
of STDP
have been explored  from many points of view (see
for example \cite{gerstner,sejnSTDP,kempterPRE,Abbott,Abarbanel,Abbott_PNAS,Witt,14,Masque} %
and papers of this special issue).
STDP has also been supposed 
to play a role in the hippocampus theta phase precession
phenomenon \cite{our_theta,11,Florian}, even though other explanations
has also been proposed for this phenomena (see \cite{theta_kempter,Leibold2008} and references
therein).
Here we analyze the role of a learning rule based on STDP in storing multiple phase-coded memories as attractor states
of the neural dynamics, and the ability of the network to selectively retrieve a stored memory, when a partial cue is presented.
The framework of storing and retrieval of memories as attractors of the dynamics is widely accepted, and recently received
experimental support, such as in the work of Wills {\em et al.} \cite{cacucci}, which gives strong experimental evidence
for the expression of rate-coded attractor states in the hippocampus.

Another characteristic of the neural network, crucial to its functioning, is its topology, that is
the average number of neurons connected to a neuron, the average length of the shortest path connecting two neurons, etc.
In the last decade, there has been a growing interest in the study of the topological structure of the brain
network \cite{sporns1,sporns2,bullmore}. This interest has been stimulated by the simultaneous development of the science of complex networks,
that studies how the behavior of complex systems (such as societies, computer networks, brains, etc.) is shaped by the way their constituent elements are connected.

A network may have the property that the degree distribution, i.e. the
probability that a randomly chosen node  is connected to $k$ other nodes, has a slow power law decay. Networks having this property
are called ``scale-free''. Barab\'asi and Albert \cite{barabasi} demonstrated that this property can originate from a process in which
each node is added preferentially to nodes that already have high degree. 
Scale-free properties have been found in functional network topology using functional magnetic resonance in human brain \cite{human},
and have been investigated in some models in relation with scale-free avalanche brain activity and criticality \cite{lucilla1,lucilla2}.

Another important class of complex networks is the so called ``small world'' networks \cite{watts}.
They combine two important properties. The first is an high level of clustering, that is an high probability of direct connection between two nodes,
given that they are both connected to a third node. 
This property usually occurs in networks where the nodes are connected preferentially
to the nearest nodes, in a physical (for example three-dimensional) space.
The second property is the shortness of paths connecting any two nodes, characteristic of random networks. 
Therefore, a measure of the small-worldness of a network is given by a high ratio of the clustering coefficient
to the path length.

There is increasing evidence that the connections of neurons in many areas of the nervous system have a small world structure
\cite{bullmore,singer,plos-cita-singer,sporns1,sporns2,hellwig}.
Up to now, the only nervous system to have been comprehensively
mapped at a cellular level is the one of Caenorhabditis elegans \cite{celegans,celegans2},
and it has been found that is has indeed a small world
structure. The same property was found for the correlation network of neurons in the visual cortex of the cat \cite{singer}.

In this paper we focus on the ability of STDP to memorize multiple phase-coded items,
in both fully connected and  sparse networks, with varying degree of small-worldness,
in a way that each phase-coded item is an attractor of the network.

Partial presentation of the pattern, i.e.\ short externally induced spike sequences,
with phases similar to the ones of the stored phase pattern,
induces the network to retrieve selectively the stored item,  as far as
the number of stored items is not larger then the network capacity.
If the network retrieves one of the stored items, the neural population 
spontaneously fires with the specific phase alignments of that pattern,
until external input does not change the state of the network. 
 
We find that the proposed learning rule  is really able to store 
multiple phase-coded patterns, and we study the network capacity, i.e.\ how many
phase-coded items can be stored and retrieved in 
the network as a function of the parameters of the network and the learning rule.

In Section \ref{sec_model} we describe the learning rule and the analog model used.
In Section \ref{sec_cap_fully} we study the case of an analog fully connected network, that is a network in which each neuron
is connected to any other neuron.
In Section \ref{sec_cap_sparse} we study instead the case of an analog sparse network, where each neuron
is connected to a finite number of other neurons, with a varying degree of small-worldness.
In Section \ref{sec_if_model} we study the case of a fully-connected 
spiking integrate-and-fire (IF) model,
and finally the summary and discussion is in Section \ref{sec_summary}.

\section{The model}
\label{sec_model}
\begin{figure}[ht]
\begin{center}
a)\includegraphics[width=5cm]{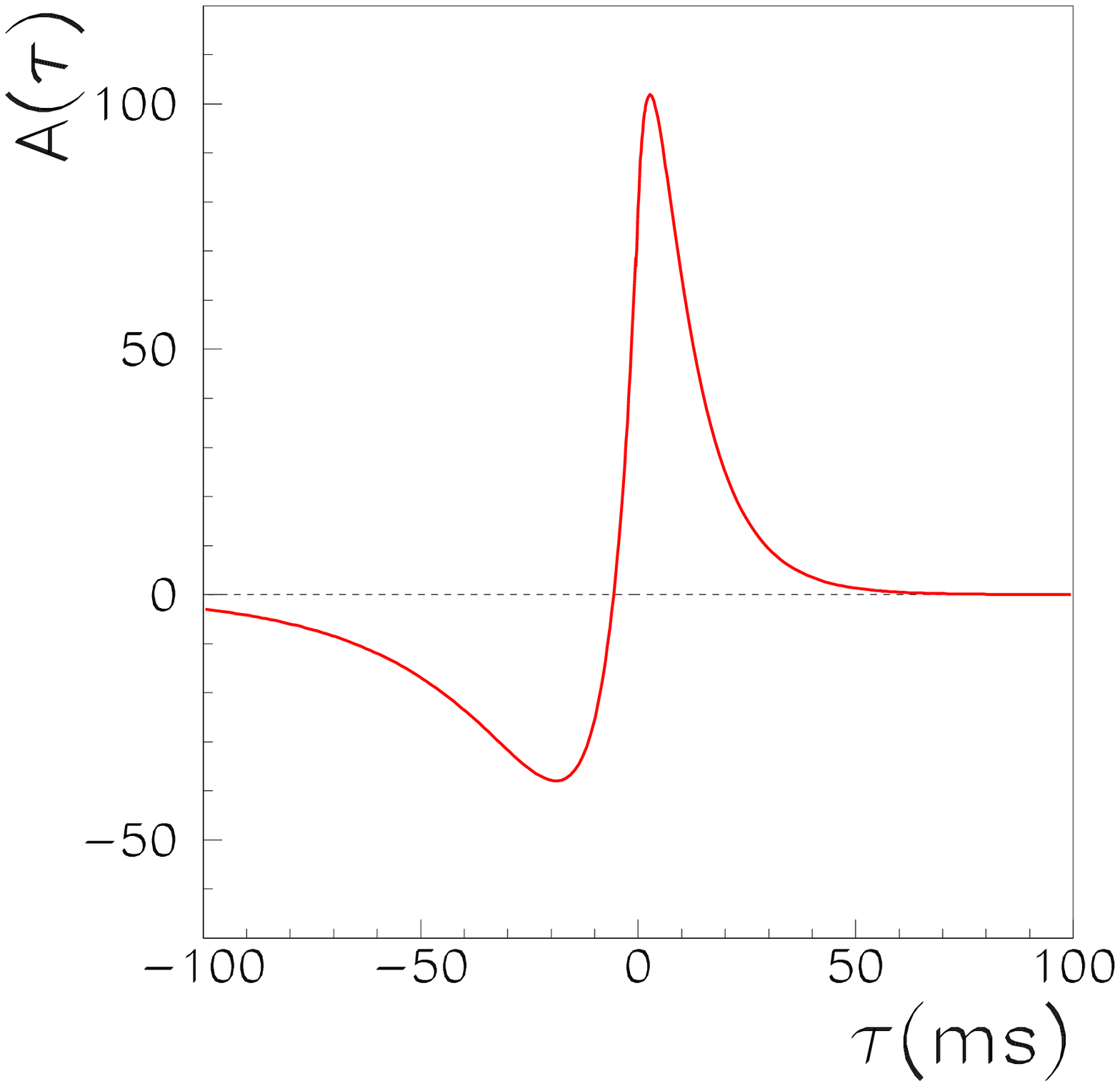}
b)\includegraphics[width=5cm]{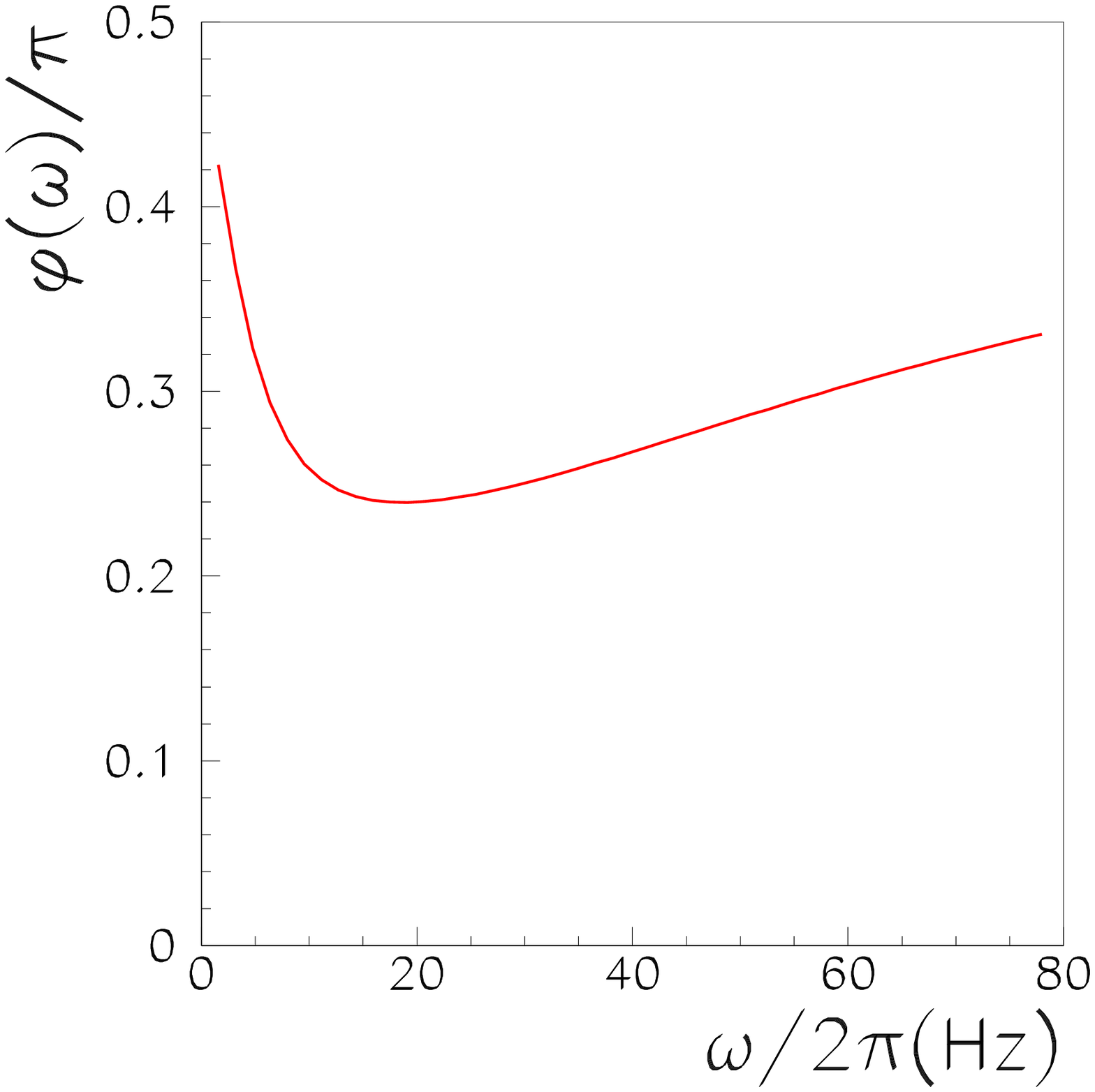}
\end{center}
\caption{%
a) The learning window $A(\tau)$ used in the learning rule in Eq.\ (\ref{lr}) to model STDP.
The window is the one introduced and motivated by \cite{Abarbanel},
$A(\tau) = a_p e^{-\tau/T_p} - a_D e^{-\eta \tau/T_p}$ if $\tau>0$,
$A(\tau) = a_p e^{\eta\tau/T_D} - a_D e^{\tau/T_D}$ if $\tau<0$,
with the same parameters used in
\cite{Abarbanel} to fit the experimental data of \cite{biandpoo},
$a_p = \gamma\,[1/T_p + \eta/T_D]^{-1}$,
$a_D = \gamma\,[\eta/T_p + 1/T_D]^{-1}$,
with $T_p=10.2$ ms, $T_D=28.6$ ms, $\eta=4$, $\gamma=42$.
Notably, this function satisfies the condition
$\int_{-\infty}^\infty A(\tau) d\tau =0$, i.e. $\tilde A(0)=0$.
b) The phase of the Fourier transform of $A(\tau)$ as a function of the frequency.
}
\label{fig_kernel}
\end{figure}

We consider a network of $N$ neurons, with $N(N-1)$ possible (directed) connections $J_{ij}$.
The synaptic connections $J_{ij}$, during the learning mode when patterns to be stored are presented, are subject to plasticity and change their efficacy according to a learning rule inspired to the STDP. 
In STDP \cite{magee, debanne, biandpoo, biandpoo2, markram, feldman} synaptic strength increases or decreases whether the presynaptic spike precedes or follows the postsynaptic one by few milliseconds,
with a degree of change that depends on the delay between pre and post-synaptic spikes,
through a temporally asymmetric learning window.
We indicate with $x_i(t)$ the activity, or firing rate, of $i$-th neuron at time $t$. It means that the firing probability of unit $i$ in the interval $(t,t+\Delta t)$ is proportional to $x_i(t)\Delta t$ in the limit $\Delta t\to 0$.
According to the learning rule we use in this work, already introduced in \cite{SZJ,NC,PREYoshioka}, the change in the connection $J_{ij}$ occurring in the time interval $[-T,0]$ can be formulated as follows:
\begin{equation}
\delta J_{ij} \propto \int\limits_{-T}^{0}\d t \int\limits_{-T}^{0}\d t^\prime \, x_i(t) A(t-t^\prime) x_j(t^\prime)
\label{lr}
\end{equation}
where $x_j(t)$ is the activity of the pre-synaptic neuron, and $x_i(t)$ the activity of the post-synaptic one.
The learning window $A(\tau)$ is the measure of the strength of synaptic
change when there is a time delay $\tau$ between pre and post-synaptic activity.
To model the experimental results of STDP, 
the  learning window $A(\tau)$ should
be an asymmetric function of $\tau$, mainly positive
(LTP) for $\tau>0$ and mainly negative (LTD) for $\tau<0$.
The shape of $A(\tau)$ strongly affects $J_{ij}$ and the dynamics of the networks, as discussed in the following.
An example of the learning window used here is shown in Fig.\ \ref{fig_kernel}.

Writing Eq.\ (\ref{lr}), implicitly we have assumed that the
effects of separate spike pairs due to STDP sum linearly.
However note that nonlinear effects have been observed when both pre- and
post-synaptic neurons fire simultaneously at more then 40 Hz \cite{15a,16a},
therefore our model holds only in the case of lower firing rates,
and  in cases where linear summation is a good approximation.

\begin{figure}[ht]
\begin{center}
a)\includegraphics[width=5cm]{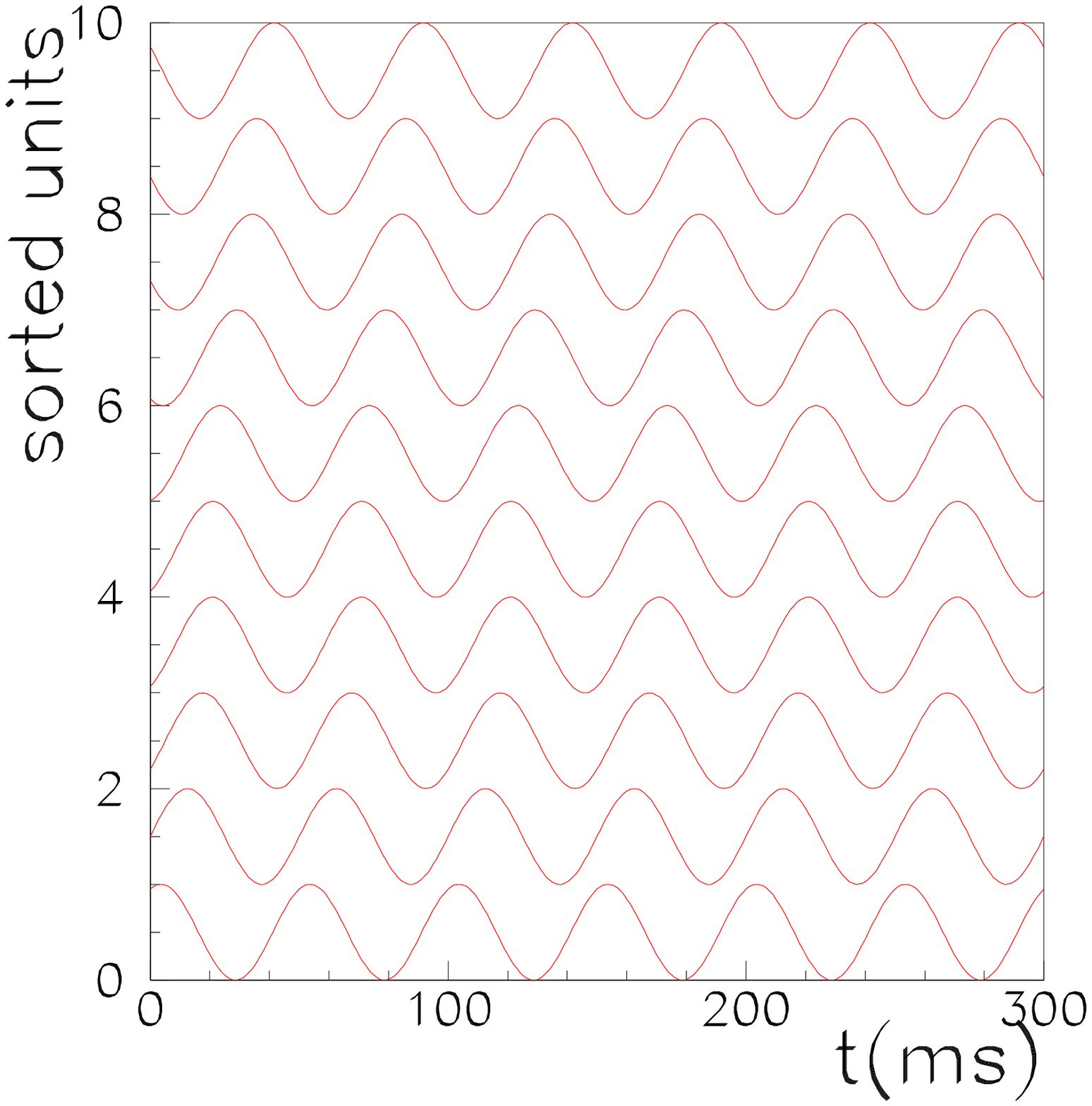}
b)\includegraphics[width=5cm]{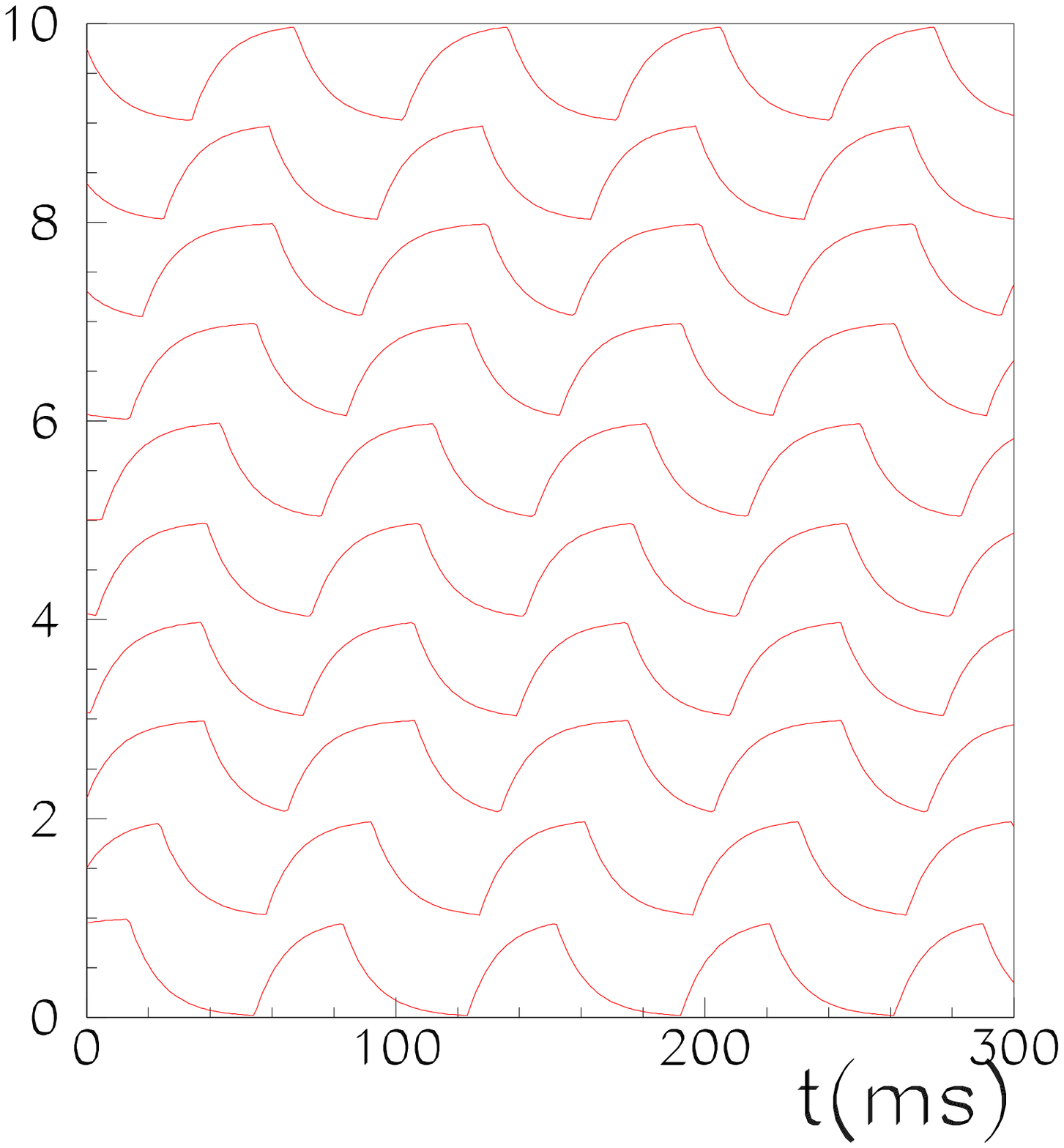}
c)\includegraphics[width=5cm]{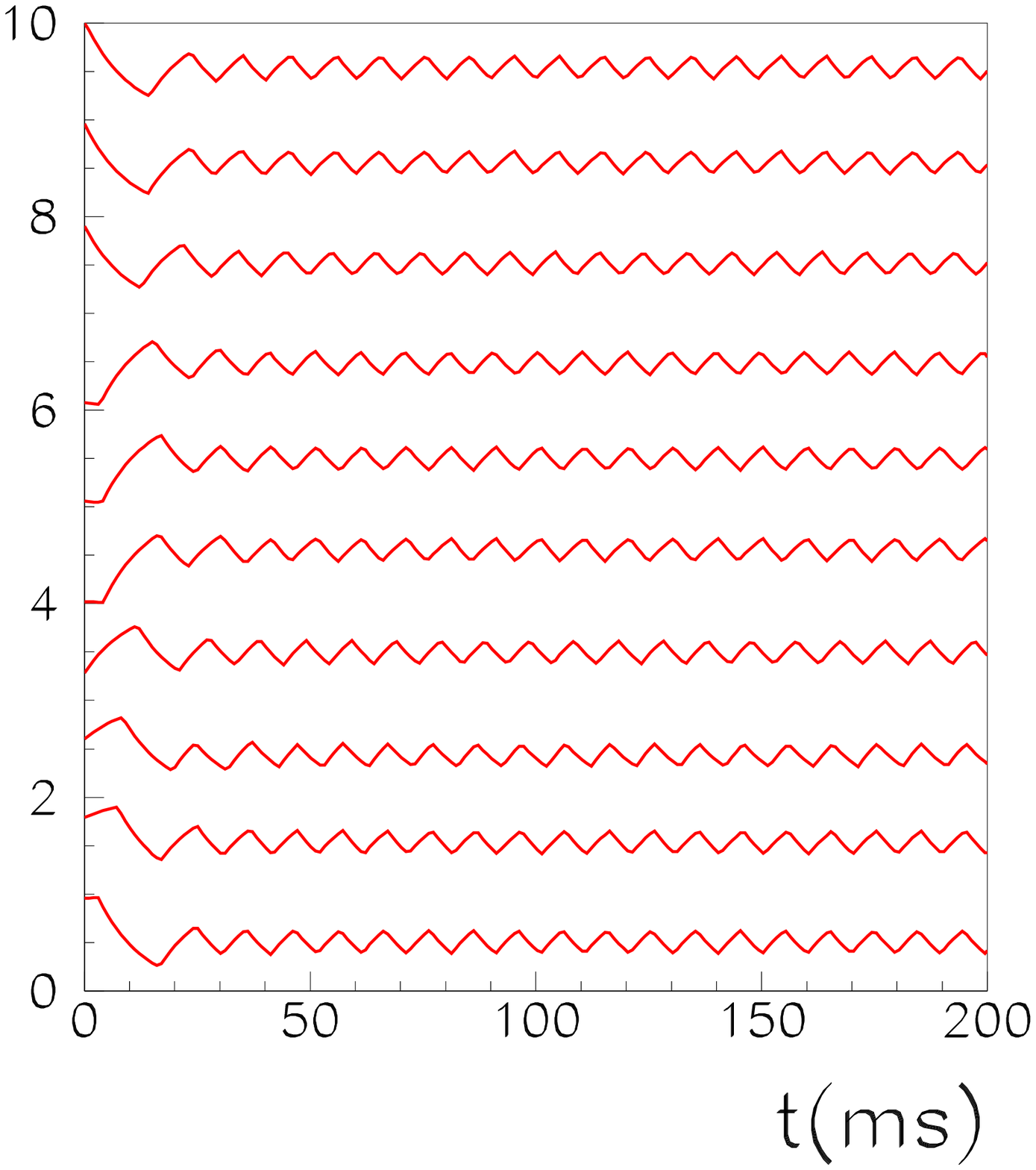}
\end{center}
\caption{%
The activity of 10 randomly chosen neurons in a network of $N=3000$ fully connected analog neurons,
with $P=30$ stored patterns.
The learning rule is given by Eq. (\ref{eq:learningrule}) with $\varphi^\ast=0.24\pi$.
Neurons are sorted by increasing phase $\phi_i^1$ of the first pattern,
and shifted correspondingly on the vertical axis.
a) The first stored pattern, that is the activity of the network given by Eq.\ (\ref{patt})
used to encode the pattern in the learning mode, with frequency $\omega_\mu/2\pi=20$ Hz.
b) The self-sustained dynamics of the network,
when the initial condition is given by the first pattern $x_i^1(0)$.
The retrieved the pattern has the same phase relationships of the encoded one. In this case the overlap
is $|m^1|\simeq 0.22$, and the output frequency is in agreement with the analytical value
$\tan(\varphi^\ast)/2\pi\tau_m=15$ Hz.
c) Same as in b), but with $\varphi^\ast=0.45\pi$. Output frequency is in agreement with the analytical value
$\tan(\varphi^\ast)/2\pi\tau_m=100$ Hz.
}
\label{fig_pattern}
\end{figure}

We consider periodic patterns of activity that  are 
periodic patterns, in which the information is encoded in the
relative phases, that is in the relative timing of the maximum firing rate of a neuron.
Therefore, we define the pattern to be stored by
\begin{equation}
x_i^\mu(t)=\frac{1}{2} \left[ 1 +\cos (\omega_\mu t - \phi^\mu_i) \right],
\label{patt}
\end{equation}
where phases $\phi^\mu_i$ are randomly chosen  from a uniform distribution in $[0,2\pi)$,
and $\omega_\mu/2\pi$ is the frequency of oscillation of the neurons
(see Fig.\ \ref{fig_pattern}a).
Each pattern $\mu$ is therefore defined  by its frequency $\omega_\mu/2\pi$, and by the specific phases
$\phi^\mu_i$ of the neurons $i=1,..,N$.

In the limit of large $T$, when the network is forced in the state given by Eq.\ (\ref{patt}),
using Eq.\ (\ref{lr}), the change in the synaptic strength will be given by
\begin{equation}
\delta J_{ij} =  \eta |\tilde A (\omega_\mu)| \cos\left[\phi_i^\mu -\phi_j^\mu-\varphi(\omega_\mu)\right] + 2\eta \tilde A(0) 
\label{JWlearning}
\end{equation}
where $\tilde{A}(\omega)$ is the Fourier transform of the kernel, defined by
\[
\tilde A (\omega)=\int_{-\infty}^{\infty} A(\tau)e^{i\omega\tau}d\tau,
\]
and $\varphi(\omega)=\arg[\tilde A (\omega)]$ is the phase of the Fourier transform.
The factor $\eta$ depends on the learning rate and on the total learning time $T$ \cite{NC,Erice}.

When we store multiple patterns  $\mu=1,2,\ldots,P$,
the learned weights are the sum of the contributions from individual patterns.
After learning P patterns, each with frequency $\omega_\mu/2\pi$ and
phase-shift $\phi_i^\mu$, we get the connections
\begin{equation}
 J_{ij}= \eta\sum_{\mu=1}^P 
|\tilde A (\omega_\mu)| \cos\left[\phi_i^\mu -\phi_j^\mu-\varphi(\omega_\mu)\right] + 2\eta P \tilde A(0).
\label{eq:learningrule}
\end{equation}
In this paper we choose $\tilde A(0)=
\int_{-\infty}^{\infty} A(\tau) d\tau =0$, which
gives a balance between potentiation and inhibition.
Notably, the condition $\tilde A(0)=0$ also holds when using the learning
window of Fig.\ \ref{fig_kernel}.
In the present study, we choose to store patterns all with the same $\omega_\mu$, 
and to ease the notation we define 
$\varphi^\ast=\varphi(\omega_\mu)$.

In the retrieval mode, the connections are fixed to the values given in Eq. (\ref{eq:learningrule}).
In the analog model, the dynamic equations for unit $x_i$  are given by
\begin{equation}
\tau_m \,\dot x_i = -x_i + F [ h_i(t) ]
\label{uno}
\end{equation}
where the transfer function $F(h)$ denotes the input-output relationship of neurons,
$h_i(t)=\sum_j J_{ij} x_j(t)$ is the local field acting on neuron $i$,
  $\tau_m $ is the time constant of neuron $i$ (for simplicity, $\tau_m$ has the same value for all neurons),
 and $J_{ij}$ is the connection after the learning procedure given in Eq.\ (\ref{eq:learningrule}).
Spontaneous activity dynamics of the coupled nonlinear system is therefore determined 
by the function $F(h)$ and  by the coupling matrix $J_ {ij}$.
We take the function $F(h)$ to be equal to the Heaviside function $\Theta(h)$. Note that in this case the learning factor $\eta$
is immaterial.

During the retrieval mode, the network selectively replays one of the stored phase-coded patterns,
depending on the initial conditions. It means that if we force the network, for $t<0$,  with an input which resembles one of the phase-coded patterns,
and then we switch off the input at times $t>0$, the network spontaneously gives sustained oscillatory activity with the relative phases of
the retrieved pattern, while the frequency can be different (see Fig. \ref{fig_pattern}).
For the analog model (\ref{uno}), where the state of the network is represented by the rates $x_i(t)$,
the same can be achieved if we simply initialize the rates at $t=0$ with values $x_i^\mu(0)$ corresponding to one of the phase-coded patterns,
or also to a partially corrupted version of it.
Analytical calculations \cite{PREYoshioka,Erice} show that the output frequency of oscillation is given by
\[
\bar \omega/2\pi=\tan(\varphi^\ast)/2\pi\tau_m,
\]
and  this is confirmed by numerical simulations
of Eq.\ (\ref{uno}) with connections given by Eq.\ (\ref{eq:learningrule}).

As an example, the learning window in Fig.\ \ref{fig_kernel},
when the frequency of the stored pattern is $\omega_\mu/2\pi=20$ Hz,
gives $\varphi^\ast=0.24\pi$,
and an output frequency of oscillation $\bar \omega/2\pi=15$ Hz (with $\tau_m= 10$ ms).
Numerical simulations of the network
with  $\varphi^\ast=0.24\pi$, $N=3000$ fully connected neurons and $P=30$ stored patterns,
are shown in Fig.\ \ref{fig_pattern}b.
In Fig.\ \ref{fig_pattern}c the case $\varphi^\ast=0.45\pi$ is shown. Here, the output frequency is much higher, while
the oscillations of the firing rates with respect to the mean value $1/2$ is much smaller.

In the following sections, we analyze the behavior when  multiple patterns are stored and we study
the network capacity as a function of learning window parameters and as a function of connectivity topology.

\section{Capacity of the fully connected network}
\label{sec_cap_fully}

In this section we study the network capacity,
in the case of fully connected network, where all the $N(N-1)$ connections are subject to the
learning process  given by Eq.\ (\ref{eq:learningrule}).

During the retrieval mode, the spontaneous dynamics of the network selectively replay one of the stored
phase-coded patterns, depending on initial condition,
so that, when retrieval is successful, the spontaneous activity of the
network is an oscillating pattern of activity with phase of firing
equal to the stored phases $\phi_i^\mu$ 
(while the frequency of oscillation is governed by the time scale of 
single neuron and by the
parameter $\varphi^\ast$ of learning window).
Similarity between the network activity during retrieval mode and the
stored phase-coded pattern is
measured by the overlap $|m^\mu|$, introduced in \cite{NC} and studied
in \cite{PREYoshioka},
\begin{equation}
|m^\mu(t)| = \left|\frac{1}{N}\sum_{j=1,\ldots,N} x_j(t) e^{i \phi_j^\mu}\right|
\label{unobis}
\end{equation}
If the activity $x_i(t)$ is equal to the pattern $x_i^\mu(t)$ in Eq.\ (\ref{patt}),
then the overlap is equal to $1/4$ (perfect retrieval),
while it is $\sim 1/\sqrt{N}$ when the
phases of firing have nothing to do with the stored phases.
Numerically we study the capacity of the network, $\alpha_c=P_{\text{max}}/N$,
where $N$ is the number of neurons and $P_{\text{max}}$ is the
maximum number of items that can be stored and retrieved successfully.

We extract $P$ different random patterns, choosing phases
$\phi^\mu_j$ randomly from a uniform distribution in $[0,2\pi)$. Then we define the connections $J_{ij}$
with the rule Eq.\ (\ref{eq:learningrule}). The values of the firing rates are initialized at time $t=0$
at the value given by Eq.\ (\ref{patt}) with $t=0$ and $\mu=1$ of the first pattern, the dynamics
in Eq.\ (\ref{uno}) is simulated, and the overlap Eq.\ (\ref{unobis}) with $\mu=1$ is evaluated.
If the absolute value $|m^\mu(t)|$ tends to a constant greater than $0.1$ at long times, then we consider
that the pattern has been encoded and replayed well by the network.
The maximum value of $P$ at which the network is able to replay the pattern is the capacity of the network.
We have verified that a small noise in the initialization do not change the results. A systematic study of
the robustness of the dynamical basins of attraction from the noise in the initialization 
has not been carried out yet.

\begin{figure}[ht]
\begin{center}
\includegraphics[width=5cm]{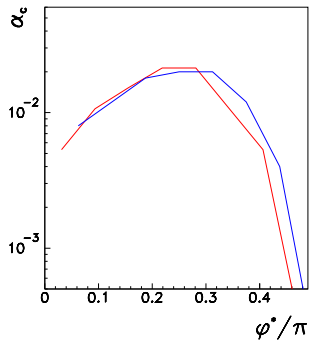}
\end{center}
\caption{%
Maximum capacity $\alpha_c=P_{\text{max}}/N$ of a network of $N=3000$
(red) and $N=6000$ (blue)
fully connected analog neurons,
as a function of the learning window asymmetry $\varphi^\ast$.
The limit $\varphi^\ast=0$ corresponds to a symmetric learning window  ($J_{ij}=J_{ji}$)
that is to output frequency $\bar\omega/2\pi$ that tends to zero.
The limit $\varphi^\ast=\pi/2$ corresponds instead to a perfectly anti-symmetric learning window,
and output frequency $\bar\omega/2\pi=\infty$.
The intermediate value $\varphi^\ast\simeq 0.25\pi$ gives the best performance of the network.
}
\label{fig_capacity_phi}
\end{figure}

Here we study the dependence of the network capacity on the learning rule parameter $\varphi^\ast$.
The parameter $\varphi^\ast$ depends on the learning window shape, and on the frequency of oscillation $\omega_\mu/2\pi$ of
the pattern presented during the learning process.
In Fig.\ \ref{fig_capacity_phi} we plot the capacity %
as a function of $0<\varphi^\ast<\pi/2$ for a fully-connected network with $N=3000$ and $N=6000$,
considering $P_{\text{max}}$ the maximum number of patterns such that
the retrieved patterns have overlaps $|m^\mu|>0.1$.
The capacity is approximately constant with the network size, showing that the maximum number of patterns
$P_{\text{max}}$ scales linearly with the number of neurons.

We see that capacity strongly depends on the
shape of learning window through parameter $\varphi^\ast$. 
The limit of $\varphi^\ast$ equal to zero corresponds to output frequency $\bar \omega/2\pi$ equal to zero,
and therefore to the limit of static output.
We see that the capacity of the oscillating network is larger then the static limit for a large range of frequencies.
When $\varphi^\ast$ approaches $\pi/2$,
then the output frequency $\bar\omega/2\pi=\tan(\varphi^\ast)/2\pi\tau_m$ tends to infinity,
and capacity decreases.
The best performance is given at intermediate values of $\varphi^\ast$.
Therefore, since $\varphi^\ast$ depends on the degree of time asymmetry of the learning window,
 we see that there is a range of time-asymmetry of the learning window which provides good capacity,
while both the case of perfectly symmetric learning window $\varphi^\ast=0$, and the case
of perfectly anti-symmetric learning window $\varphi^\ast=\pi/2$, give worse
capacity performances.
Interestingly, the learning window in Fig.\ \ref{fig_kernel} gives intermediate values
of $\varphi^\ast$ for a large interval of frequencies $\omega_\mu$.

Note that the decrease in the capacity of the network when the phase $\varphi^\ast$ approaches $\pi/2$
is essentially due to the fact that the oscillations of the firing rates with respect to the mean value $1/2$
become small in this regime.
When the firing rates tend to a constant, the overlap defined by Eq.\ (\ref{unobis}) goes to zero.

\section{Capacity of the sparse network}
\label{sec_cap_sparse}

In this section we study the capacity of the network described through Eq. (\ref{uno}) in the case of sparse connectivity,
where only a fraction of the connections are subject to the
learning rule given by Eq.\ (\ref{eq:learningrule}), while all the others are set to zero.
The role of connectivity's topology  is also investigated.
We start from a network in which neurons are put on the vertices of a two or three dimensional lattice; each neuron is connected only to neurons within a given distance (in units of lattice spacings)
and we call $z$ the number of connections of a single neuron.
For each neuron, we
then ``rewire'' a finite fraction $\gamma$ of its connections,
deleting the existing short range connections and creating, in place of them,  long range
connections to  randomly chosen neurons.
\begin{figure}[ht]
\begin{center}
a)\includegraphics[width=5cm]{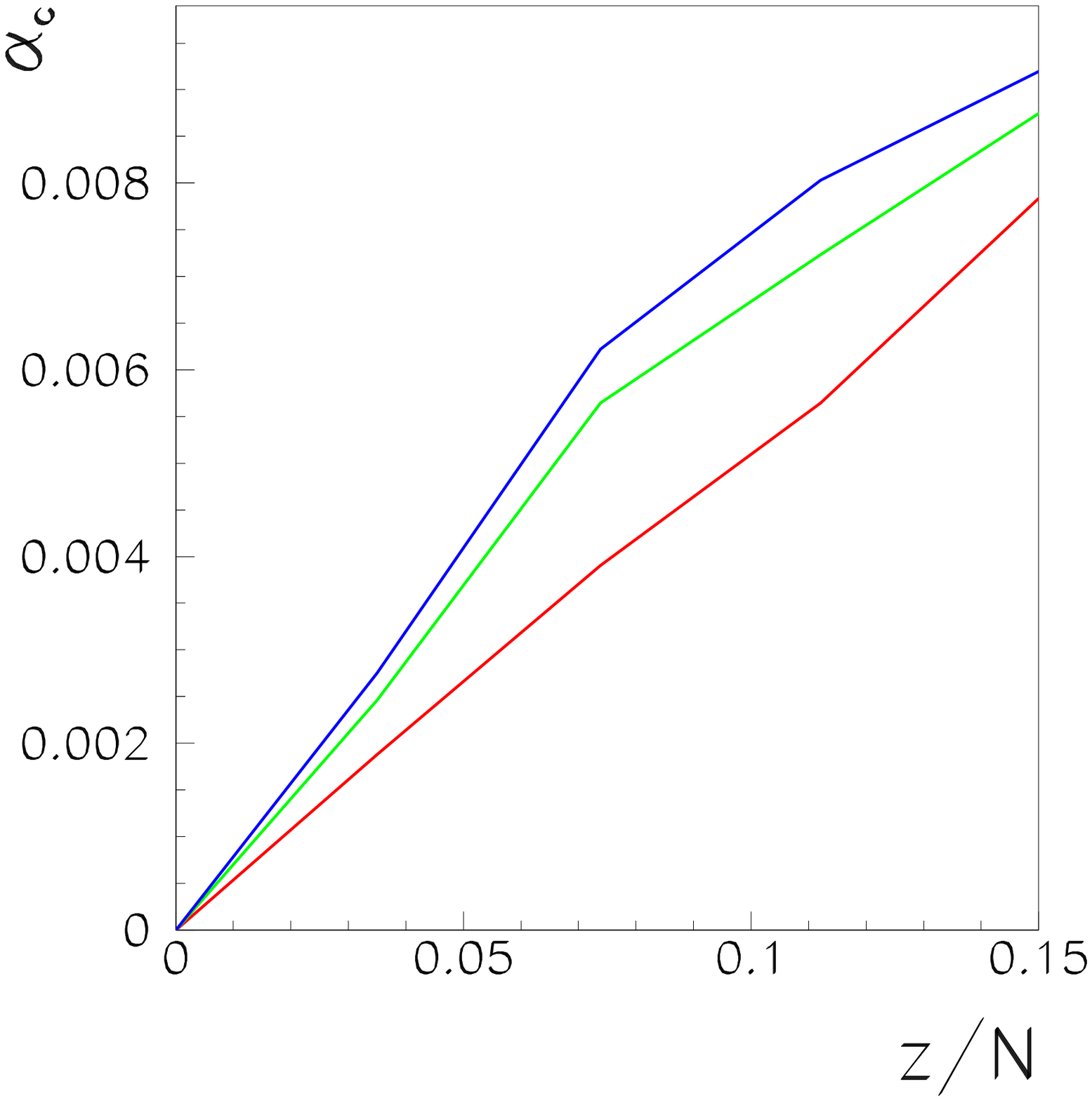}
b)\includegraphics[width=5cm]{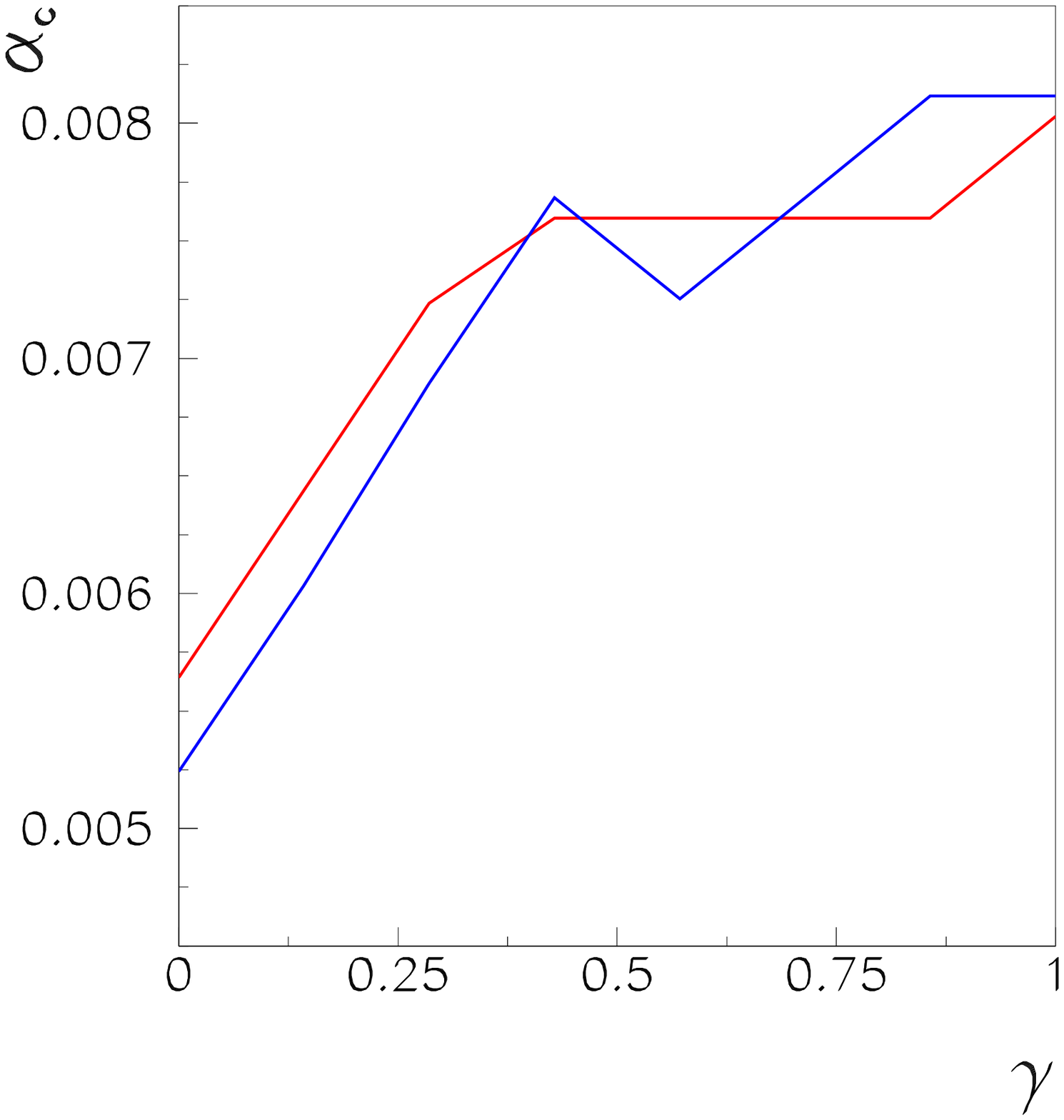}
c)\includegraphics[width=5cm]{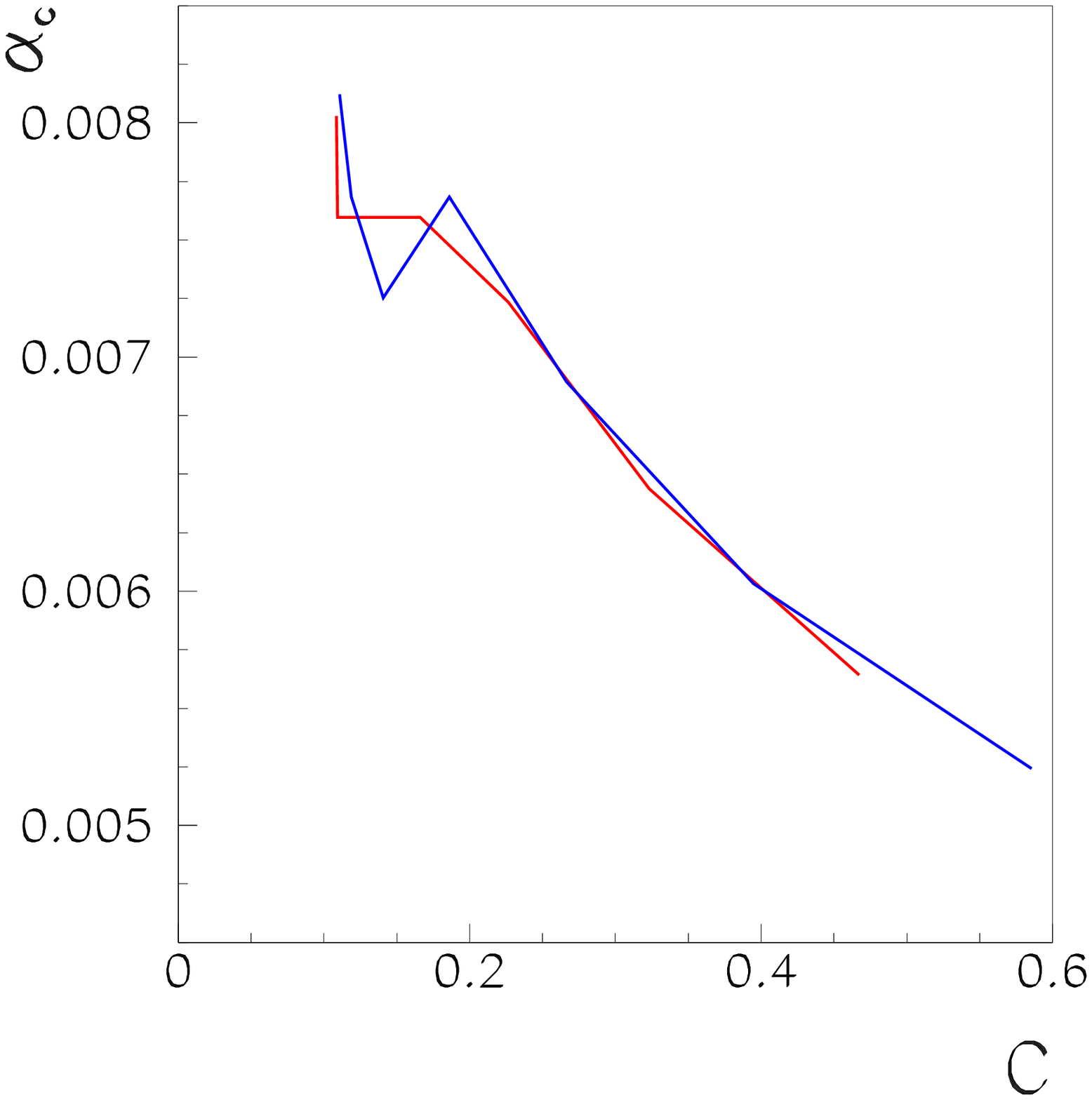}
\end{center}
\caption{%
a) Maximum capacity $\alpha_c=P_{\text{max}}/N$ for a network with $24^3$ analog neurons,
with $z$ connections per neuron, with $\varphi^\ast$
fixed to its optimal value $0.24\pi$.
The red curve corresponds to $\gamma=0$, that is to a network with only short range connections.
The green one to $\gamma=0.3$, and the blue one to $\gamma=1$, that is to a random network where the
finite dimensional topology is completely lost.
b) Maximum capacity $\alpha_c=P_{\text{max}}/N$, for the same values of $N$ and $\varphi^\ast$ and for
connectivity $z/N=0.11$, as a function of $\gamma$, for the $24^3$ lattice (red) and for a two dimensional
$118^2$ lattice (blue).
c) Same data as in b), but as a function of the clustering coefficient $C$.
}
\label{fig_sparse}
\end{figure}

We consider a three-dimensional network with $24^3$ neurons, and a two-dimensional one with $118^2$ neurons.
In Fig.\ \ref{fig_sparse}a we plot the maximum capacity $\alpha_c=P_{\text{max}}/N$
as a function of the connectivity $z/N$ for three different values of $\gamma$, for the three-dimensional case.
The value $\gamma=0$ (red curve) corresponds to the pure short range network,
in which all connections are between neurons within a given distance
on the three-dimensional lattice, $\gamma=1$ (blue curve) corresponds to the random network,
where the three-dimensional topology is completely lost,
and $\gamma=0.3$ to an intermediate case,
where 30\% of the connections are long range, and the others are short range.

Considering that the capacity of the fully connected network ($z/N=1$) is $\alpha_c\simeq 0.02$,
we see that the random network with $z/N\simeq 0.1$ already has $\sim 40\%$
of the capacity of the fully connected network.
This means the capacity does not scale linearly with the density of connections, and when
the density of connections grows there is a sort of saturation effect, due to the presence of
partially redundant connections.

Then, we look at the dependence of the capacity from the fraction of short range and long range
connections for $z/N\simeq 0.11$. In Fig.\ \ref{fig_sparse}b,
we see that the capacity gain, with respect to the short range network, given by $30\%$ long range connections
($\gamma=0.3$), is about $80\%$ the one of the fully random network ($\gamma=1$).
Therefore, the presence of a small number of long range connections is able to amplify
the capacity of the network. This is shown in Fig.\ \ref{fig_sparse}b,
where the capacity as a function of $\gamma$ for $z/N=0.11$ is plotted (red curve).
Note that the above effect is not so important in smaller networks,
for example with $N=18^3$ neurons, where the capacity of the short range network
is nearer to that of a random network.
Therefore, it seems plausible that, for very large networks,
the amplifying effect of a small fraction of long range connections will be even stronger.

It is reasonable to suppose that the fraction of long range connections, in real networks,
is determined by a trade-off between the increase of capacity given by long range connections,
and their higher wiring cost. A large amplifying effect of long range connections on the capacity,
together with a large wiring cost with respect to short range ones,
will result in the optimal topology of the network being small-world like, with
a small fraction of long range connections, as observed in many areas of the brain,
from C. elegans \cite{celegans,celegans2} to the visual cortex of the cat \cite{singer}.

The experiments on a two-dimensional lattice with $118^2$ neurons do not show any qualitative
differences with respect to the three-dimensional case. In Fig.\  \ref{fig_sparse}b
we plot the capacity of the two-dimensional network (blue curve) as a function of $\gamma$,
along with the capacity of the three-dimensional one (red curve).

The Fig.\ \ref{fig_sparse}c shows the same data of Fig.\ \ref{fig_sparse}b,
but as a function of the clustering coefficient $C$,
defined as the probability that two sites, neighbors of a given site, are neighbors themselves.
As reported in Fig.\ \ref{fig_sparse}c,
once fixed the number of connections per neuron $z$, 
the lower the clustering coefficient, the higher the capacity of the network.
This is reasonable, because a high clustering coefficient means that connections
will be partially redundant, as already observed in the case of the dependence on the connectivity $z/N$.

Note that the mean path length $\lambda$, for the considered value of $z/N=0.11$, is a decreasing function of $\gamma$ from $\gamma=0$ up to
$\gamma=0.01$, and then remains practically constant for higher values of $\gamma$.
This means that the ratio between $C$ and $\lambda$, that is the small-worldness of the network, has a maximum about $\gamma=0.01$.
The capacity therefore is not an increasing function of the small-worldness. Only when the wiring cost of the connections
is taken in account, the optimal topology turns out to be one with a small fraction of long range connections.
Note also that for more realistic values of $z/N$, much lower than $z/N=0.11$, the maximum of the small-worldness shifts to
higher values of $\gamma$.


\begin{figure}[ht]
\begin{center}
a)\includegraphics[width=5cm]{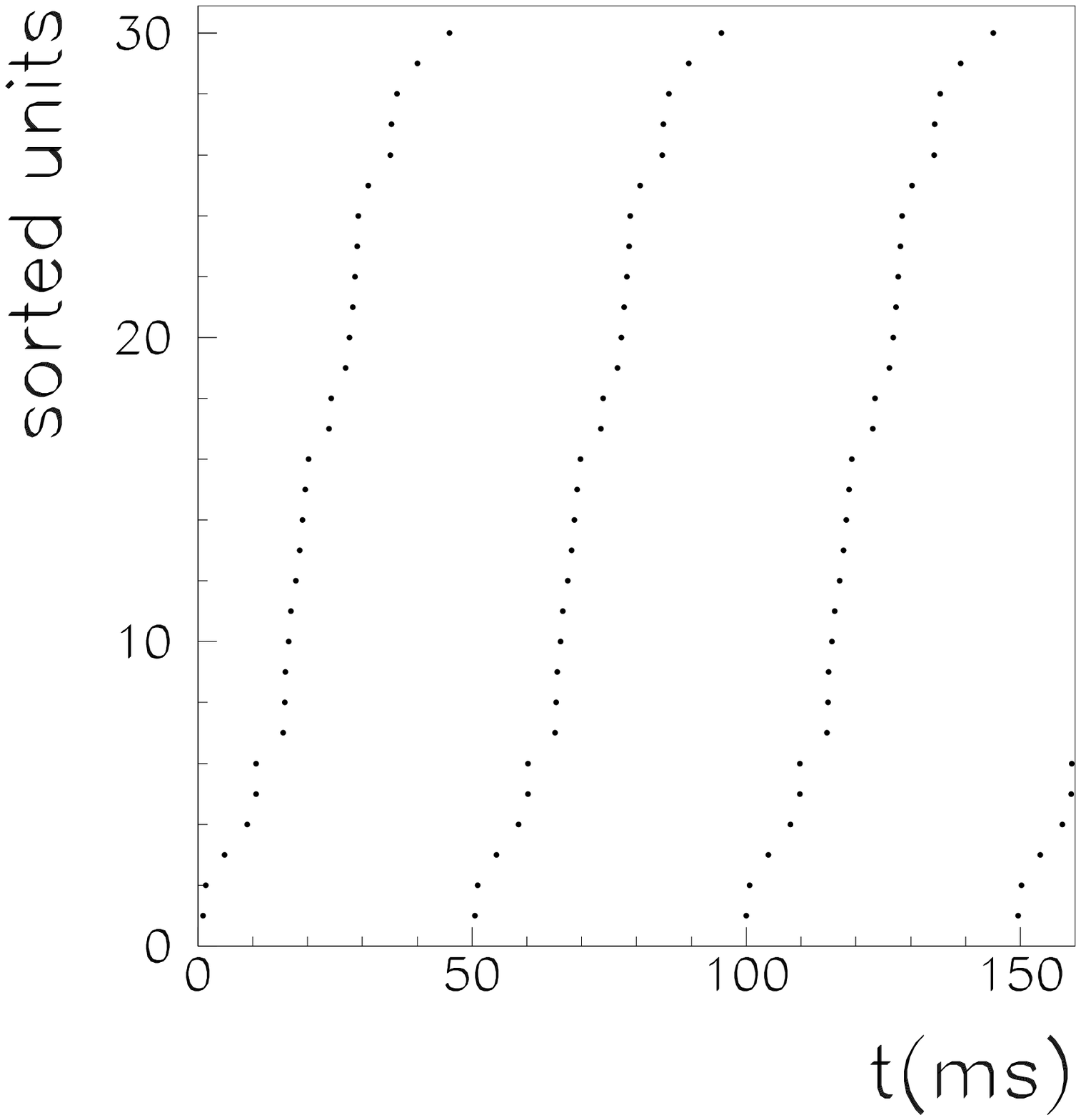}\\
b)\includegraphics[width=5cm]{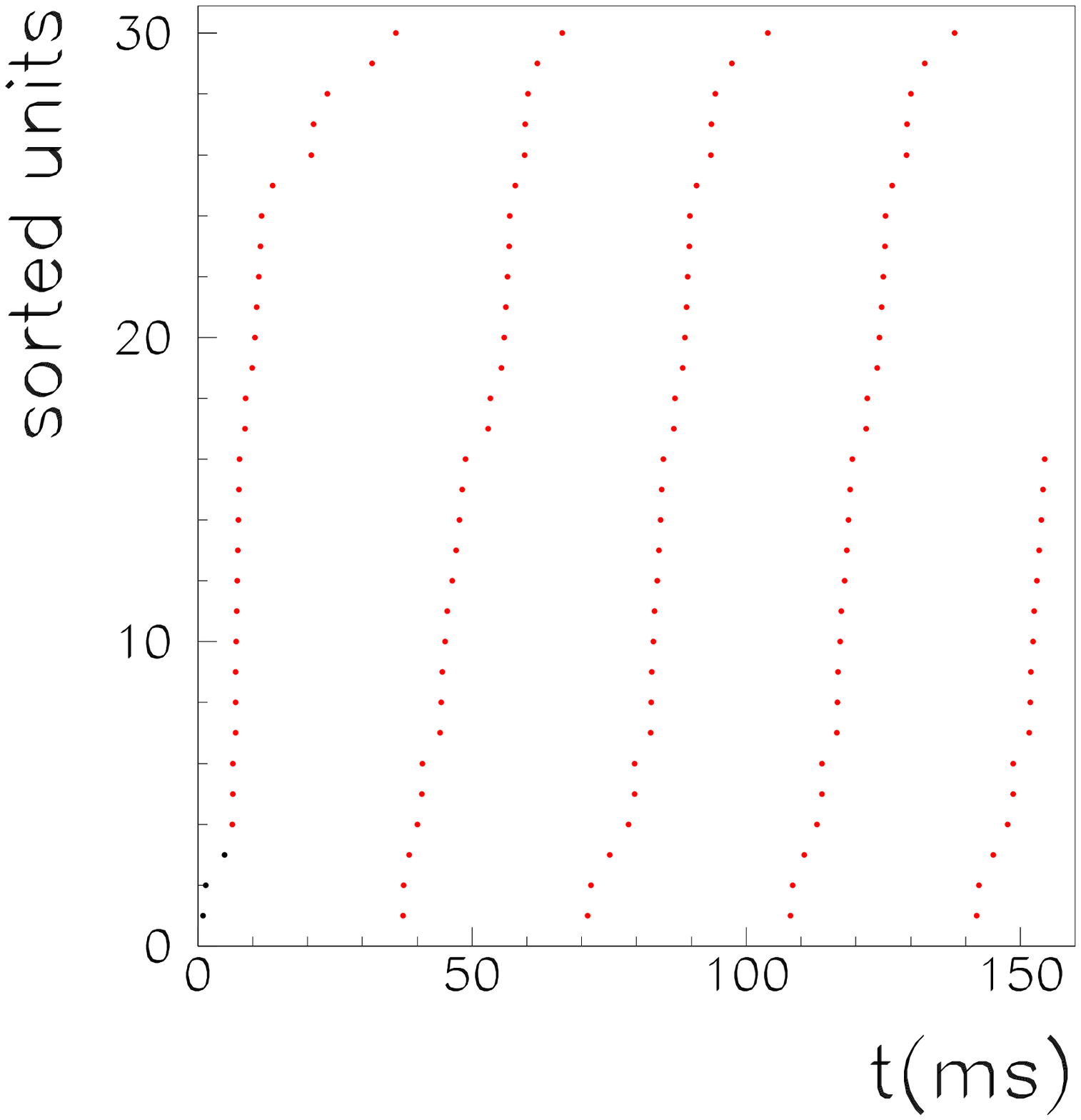}
c)\includegraphics[width=5cm]{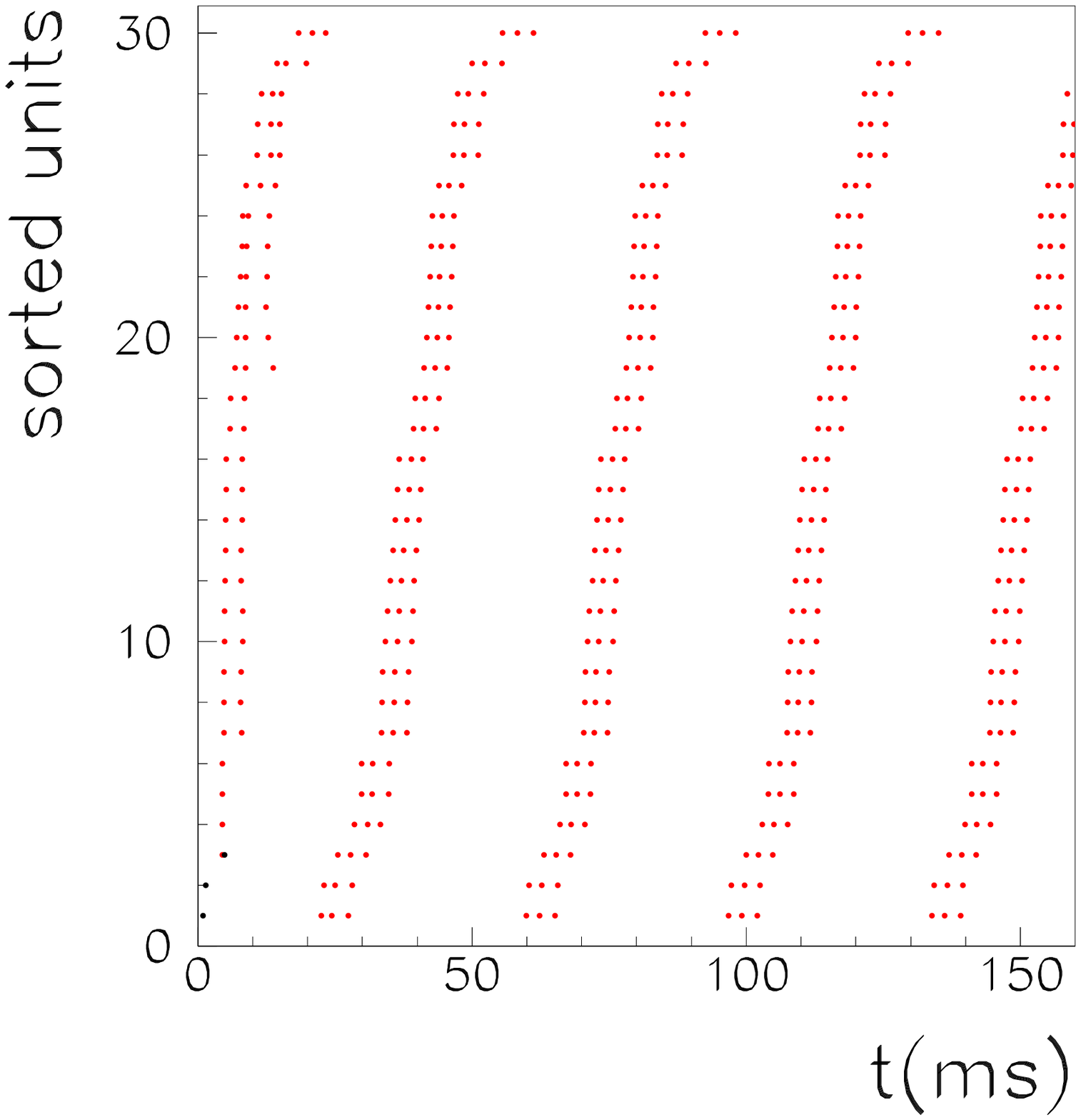}
d)\includegraphics[width=5cm]{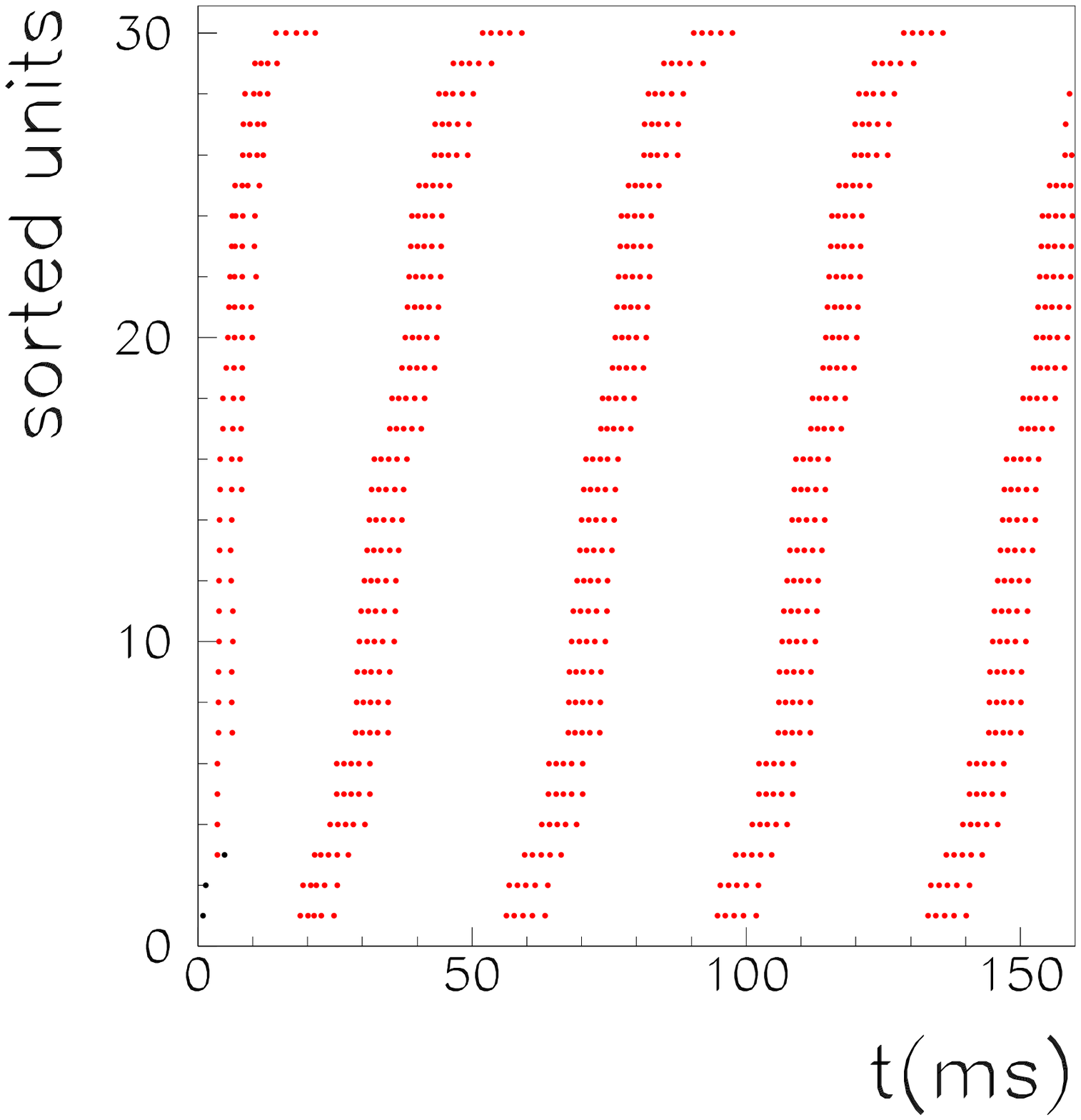}
\end{center}
\caption{%
Recall of a pattern by the spiking IF model, with $N=1000$ neurons and $\varphi^\ast=0.24\pi$.
One pattern ($P=1$) defined by the phases $\phi_i^\mu$ is stored in the network with the rule given by Eq.\ (\ref{eq:learningrule}).
Then a short train of $M=150$ spikes,
at times $t_i=\phi_i^\mu/\omega_\mu$ with $\omega_\mu/2\pi=20$ Hz, is induced on the neurons that have the $M$ lowest phases $\phi_i^\mu$.
This short train triggers the replay of the pattern by the network.
Depending on the value of $T$, the phase-coded pattern is replayed with a different number of spikes per cycle.
Note that changing $T$ in our model may correspond to 
a change in the value of physical threshold or in the value of
the parameter $\eta \tilde A(\omega_\mu)$ appearing in the synaptic connections $J_{ij}$ learning rule.
a) 30 neurons of the network are randomly chosen and sorted by the value of the phase $\phi_i^\mu$. Then
the phases of the encoded pattern are shown,
plotting on the $x$ axis the times $(\phi_i^\mu+2\pi n)/\omega_\mu$, and on the $y$ axis the label of the neuron.
b) The replayed pattern with threshold $T=85$ is shown, plotting on the $x$ axis the times of the spikes, and on the $y$ axis the label of the spiking neuron.
Black dots represent externally induced spikes, while red dots represent spikes generated by the intrinsic dynamics of the network.
c) The replayed pattern with threshold $T=50$. d) The replayed pattern with threshold $T=35$.
}
\label{fig_T}
\end{figure}
\section{Integrate and fire model}
\label{sec_if_model}
The previous results refer to network dynamics described in Eq.\ (\ref{uno}),
that is simple enough to admit analytical predictions for dynamics when the connectivity is
given by Eq.\ (\ref{eq:learningrule}).
The simple model defined by Eq.\ (\ref{uno})  has state variables $x_i(t)$, representing instantaneous firing rate or probability of firing,
 and it has  only one time scale $\tau_m$, which is the time constant of a single unit,
allowing us to focus on the effect of the learning window shape and connectivity structure.
However, we expect that, while details of dynamics may depend on the model of single unit, the  crucial results of storing and recall of phase-coded patterns %
can be seen also in a spiking model. Therefore we simulate a Leaky Integrate and Fire (IF) spiking model.
We use a simple Spike-Response-Model formulation (SRM)
\cite{bookG,SRM}  of the leaky Integrate and Fire model.
While integrate-and-fire models are usually defined in terms of
differential equations, the SRM expresses the membrane potential at
time t as an integral over the past \cite{bookG}. This allows us to
use an event-driven programming and makes the numerical simulations faster than in
the differential equation formulation.

\begin{figure}[ht]
\begin{center}
a)
\includegraphics[width=5cm]{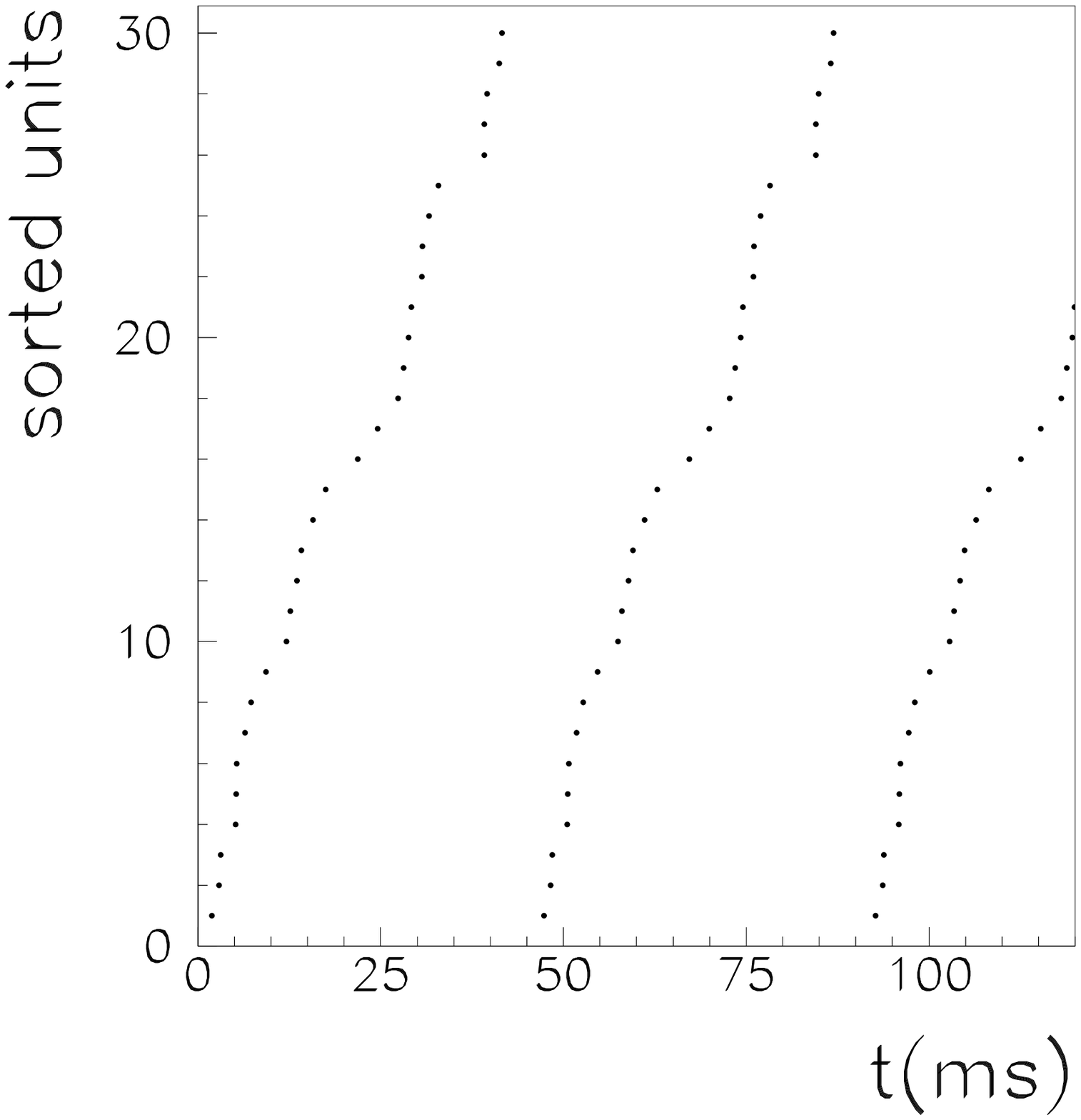}
b)%
\includegraphics[width=5cm]{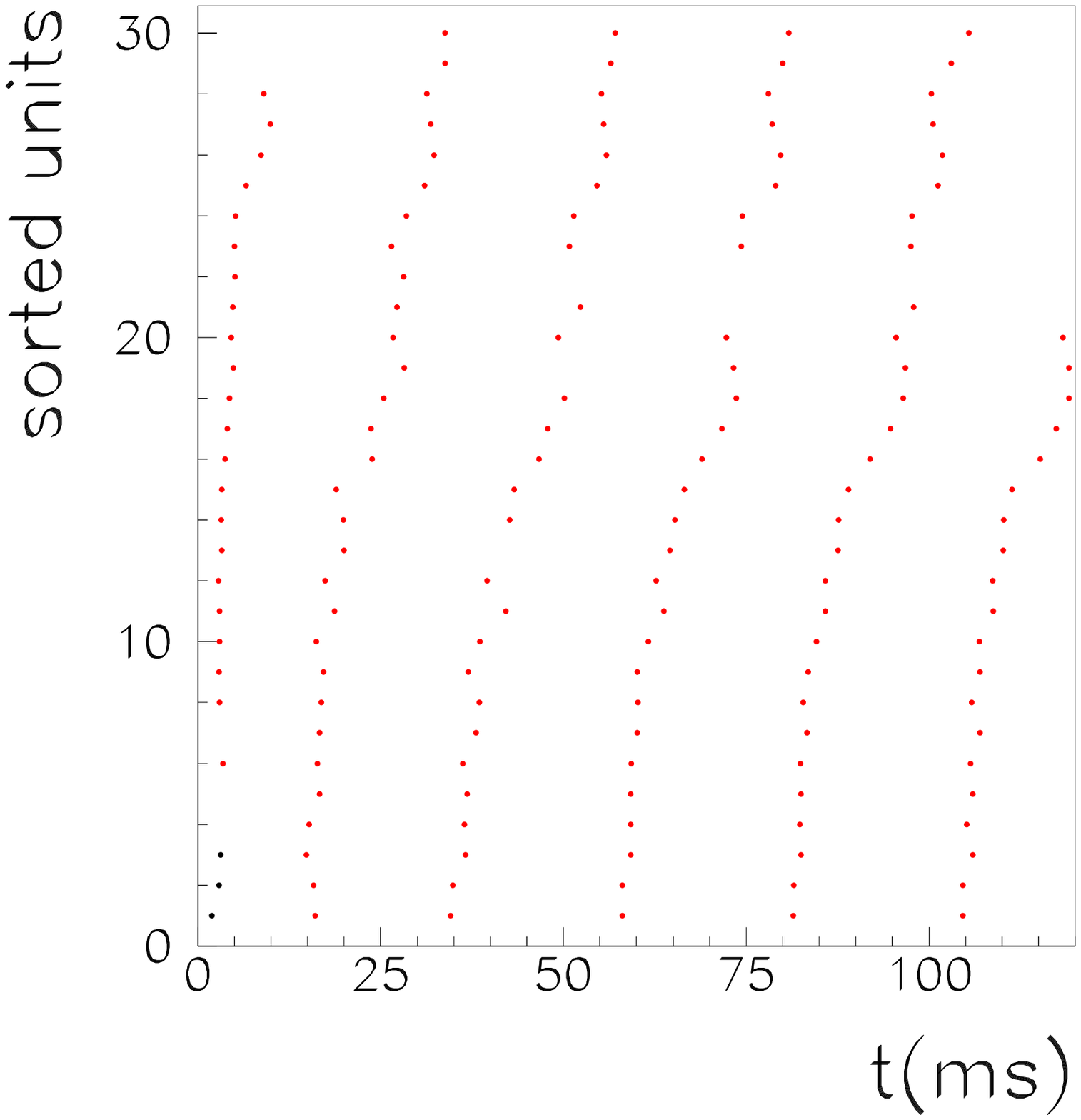}
c)%
\includegraphics[width=5cm]{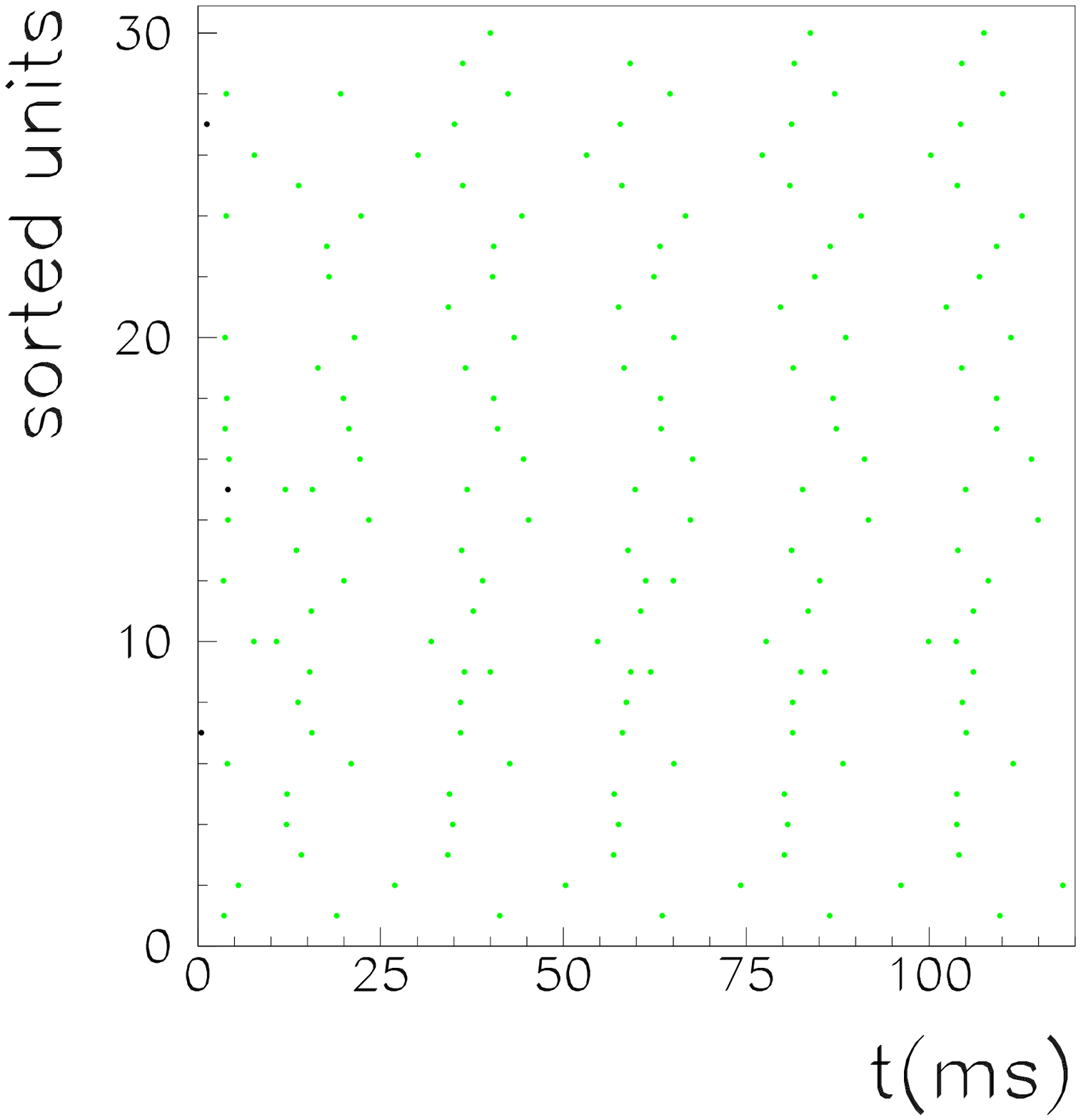}
\\
d)%
\includegraphics[width=5cm]{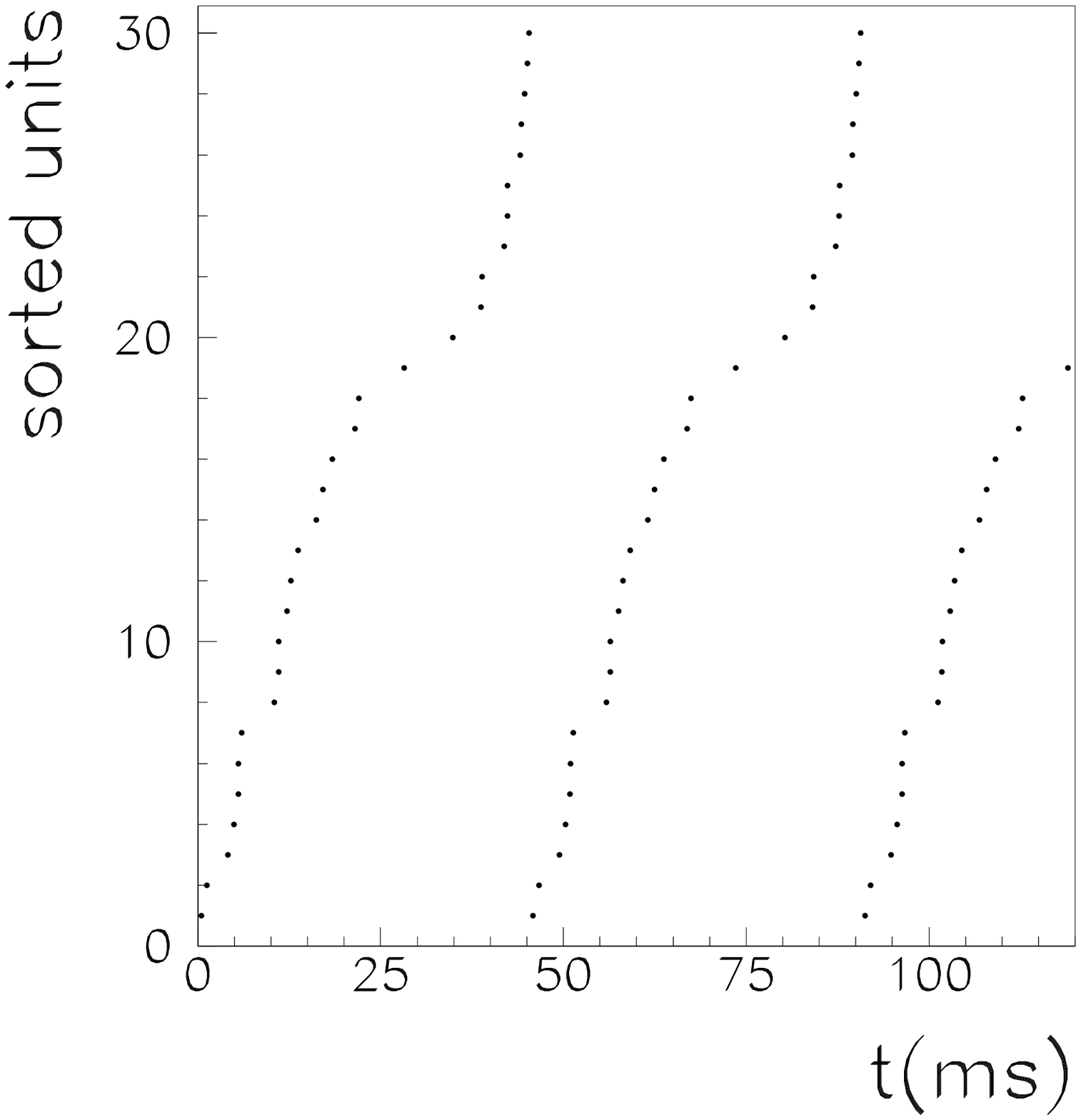}
e)%
\includegraphics[width=5cm]{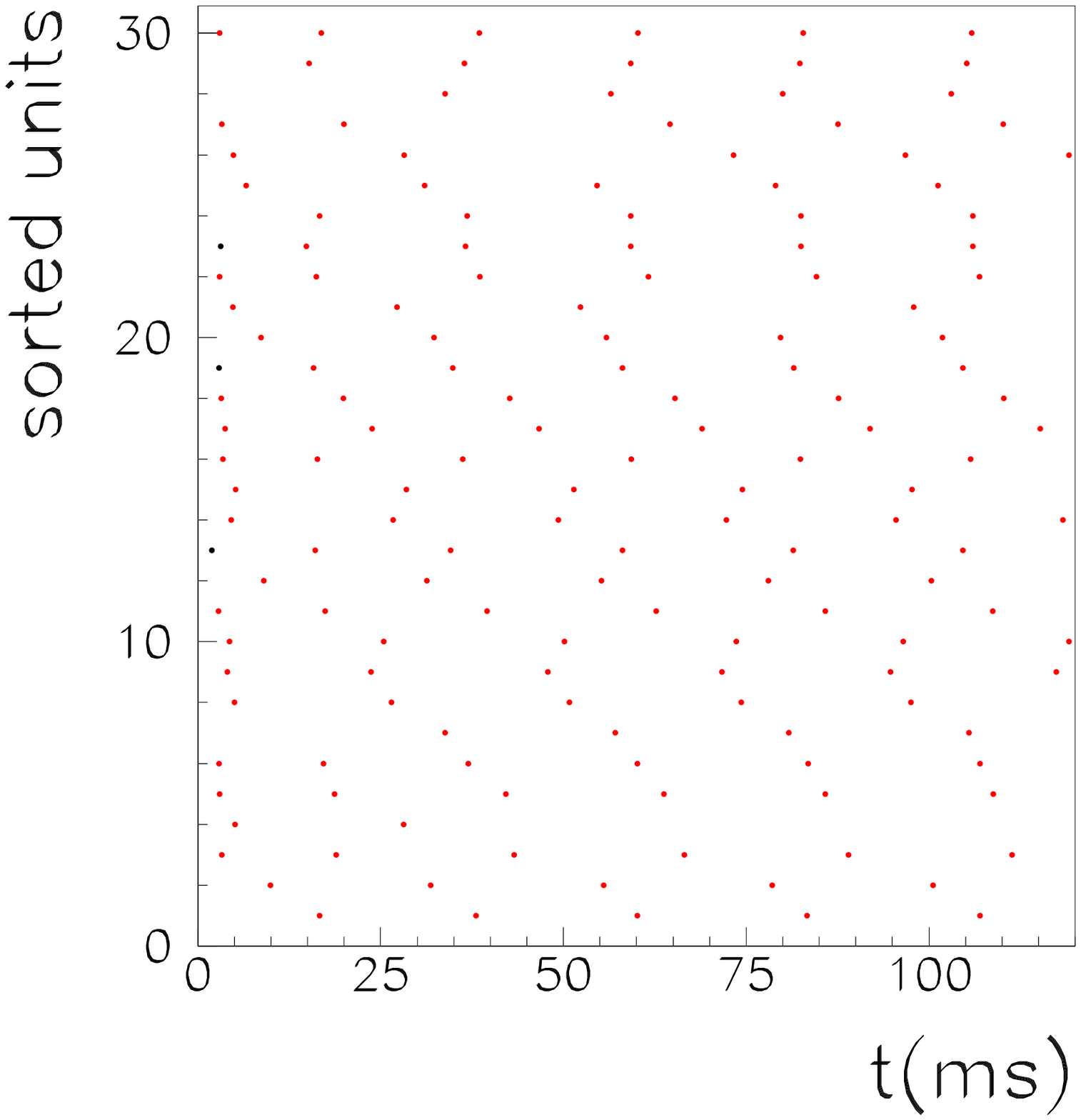}
f)%
\includegraphics[width=5cm]{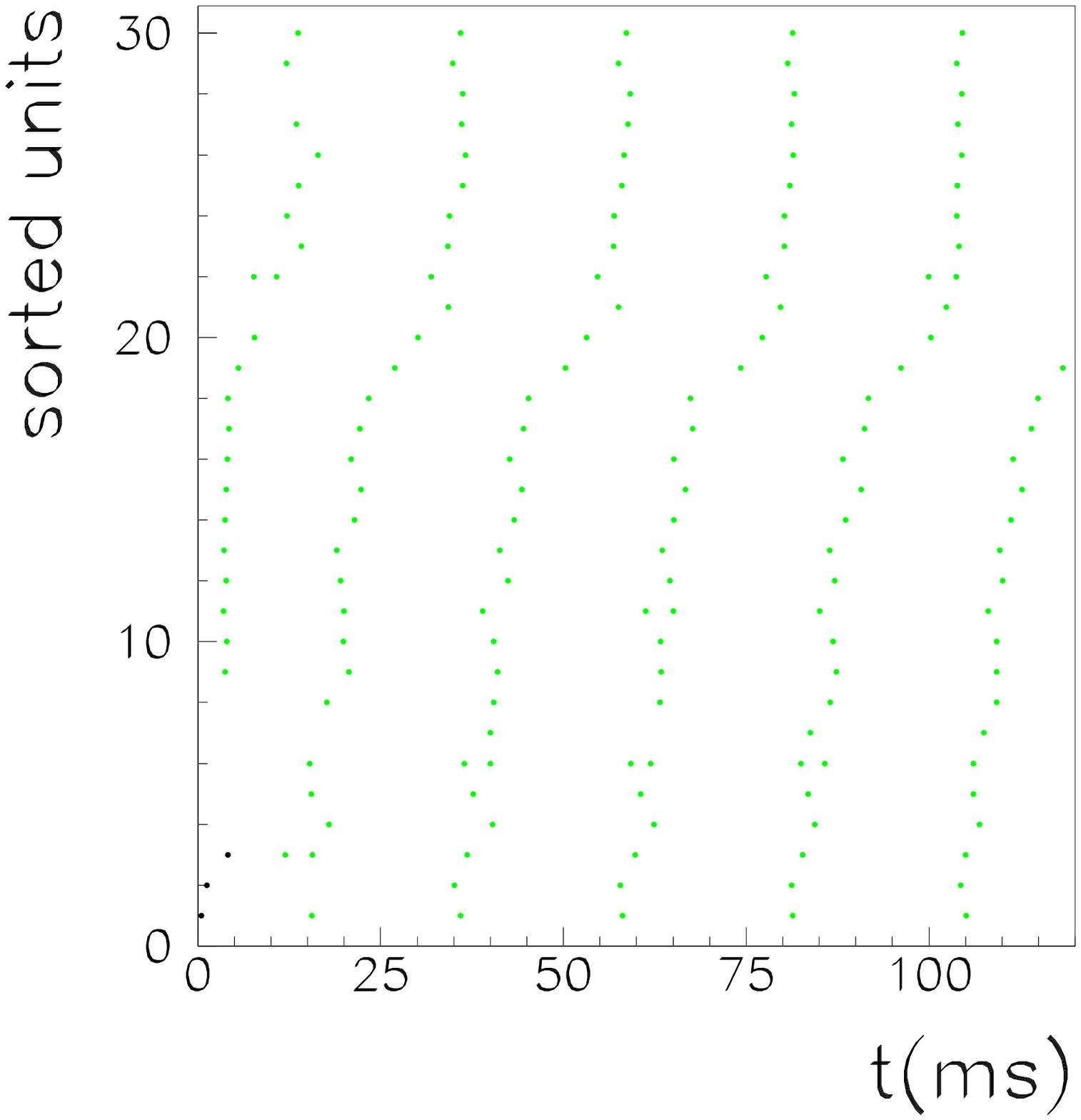}
\end{center}
\caption{%
Example of selective retrieval of different patterns, in a network with $N=1000$ IF neurons and $P=5$ stored patterns.
Here we choose the asymmetry parameter to be $\varphi^\ast=-0.4\pi$.
30 neurons are chosen randomly, and then sorted by the value of $\phi_i^1$ in (a-c), and sorted by the value of $\phi_i^2$
in (d-f). In a) and d) we show the phases of the first two stored patterns,
plotting the times $(\phi_i^\mu+2\pi n)/\omega_\mu$ respectively for $\mu=1$ and $\mu=2$.
In b) and e) we show the generated dynamics when a short train of $M=100$ spikes corresponding to the $\mu=1$ pattern is
induced on the network, while in c) and f) the dynamics when the $\mu=2$ pattern is instead triggered. The network dynamics selectively replay the stored pattern, depending on the partial cue stimulation.
Note that, because the parameter $\varphi^\ast$ here is greater than in Fig.\ \ref{fig_T},
the frequency of the retrieved pattern here is greater.
}
\label{fig_varphi}
\end{figure}

Each presynaptic spike $j$, with arrival time $t_j$, is supposed to add to the
membrane potential a postsynaptic potential of the form $J_{ij} \epsilon(t-t_j)$, where
\begin{equation}
\epsilon(t-t_j)= K \left[\exp\left(-\frac{t-t_j}{\tau_m}\right) - \exp\left(-\frac{t-t_j}{\tau_s}\right) \right] \Theta(t-t_j)
\label{tre}
\end{equation}
where $\tau_m$ is the membrane time constant (here 10 ms), $\tau_s$ is the synapse time
constant (here 5 ms), $\Theta(t)$ is the Heaviside step function,
and K is a multiplicative constant chosen so that the maximum value
of the kernel is 1. The sign of the synaptic connection $J_{ij}$ set the sign of the postsynaptic potential change.
The synaptic signals received by neuron $i$, after time $t_i$ of the last spike of neuron $i$, are added
to find the total postsynaptic potential
\begin{equation}
h_i(t)= \sum_{j/t_j>t_i} J_{ij} \epsilon(t-t_j).
\label{quattro}
\end{equation}
When the postsynaptic potential of neuron $i$ reaches the threshold $T$, a postsynaptic 
spike is scheduled, and postsynaptic potential is set to the resting value zero.
We simulate this simple model with $J_{ij}$ taken from the learning rule given by Eq.\ (\ref{eq:learningrule}),
with $P$ patterns in a network of $N$ units.
The learning rule Eq.\ (\ref{eq:learningrule}) comes out of the learning process given
by Eq.\ (\ref{lr}),
when a sequence of spikes is learned, and spikes are generated in such
a way that the probability that unit $i$ has a spike
in the interval $(t,t+\Delta t)$ is proportional to $x_i^\mu(t)\Delta t$
in the limit $\Delta t\to 0$,
with the rate $x_i^\mu(t)$ given by Eq.\ (\ref{patt}).
The $J_{ij}$ are measured in units such that $\eta \tilde A(\omega_\mu)= 1$.

After the learning process, to recall one of the encoded patterns,
we give an initial signal made up of $M<N$ spikes, taken from the stored pattern $\mu$,
and we check that after this short signal the spontaneous dynamics of the network 
gives sustained activity with spikes aligned to the phases $\phi_i^\mu$ of pattern $\mu$ (Fig.\ \ref{fig_T}).

We also investigate the role of the threshold T.
As shown in Fig.\ \ref{fig_T}, when the threshold T is lower a burst of activity takes place within each cycle, with phases aligned with the pattern.  Therefore the same phase-coded pattern is retrieved, but with a different number of spikes per cycle. 
This open the possibility to
have a coding scheme in which the phases encode pattern's informations, and
rate in each cycle represents the strength and saliency of the retrieval or it may encode another variable.
The recall of the same phase-coded pattern with different
number of spikes per cycle, shown in Fig.\ \ref{fig_T},
accords well with recent observation of Huxter {\em et al.} \cite{huxter_burgess}
in hippocampal place cells, showing occurrence of the same phases with
different rates. They show that the phase of firing and firing rate are
dissociable and can represent two independent variables, e.g. the animal’s location within
the place field and its speed of movement through the field.
Note that a change of $T$ in our model may correspond to 
a change in the value of physical threshold, or to a change in the value of
the parameter $\eta \tilde A(\omega_\mu)$ appearing in
the synaptic connections $J_{ij}$.

The role of parameter $\varphi^\ast$ in connections $J_{ij}$
in Eq. (\ref{eq:learningrule}) is that of changing the output frequency of
collective oscillation during retrieval. Here, as well as in the
analog model of Eq. (\ref{uno}), lower values of $\varphi^\ast$ correspond to
lower frequencies, even though in the spiking model
both time constants of the single neuron ($\tau_m$ and $\tau_s$) play a role to set output frequency.
Therefore, the simple formula that holds for
the output frequency of the  analog model is not valid in the spiking model.
Output activity during recall, for $\varphi^\ast = 0.24\pi$, is shown
in Fig.\ \ref{fig_T}, while output activity when $\varphi^\ast = 0.4\pi$ has
a higher frequency, as shown in Fig. \ref{fig_varphi}.
Selective recall of two of the stored patterns are shown in Fig. \ref{fig_varphi}
 when  $P=5$ patterns are stored, and $\varphi^\ast = 0.4\pi$. 

\begin{figure}[ht]
\begin{center}
a)
\includegraphics[width=5cm]{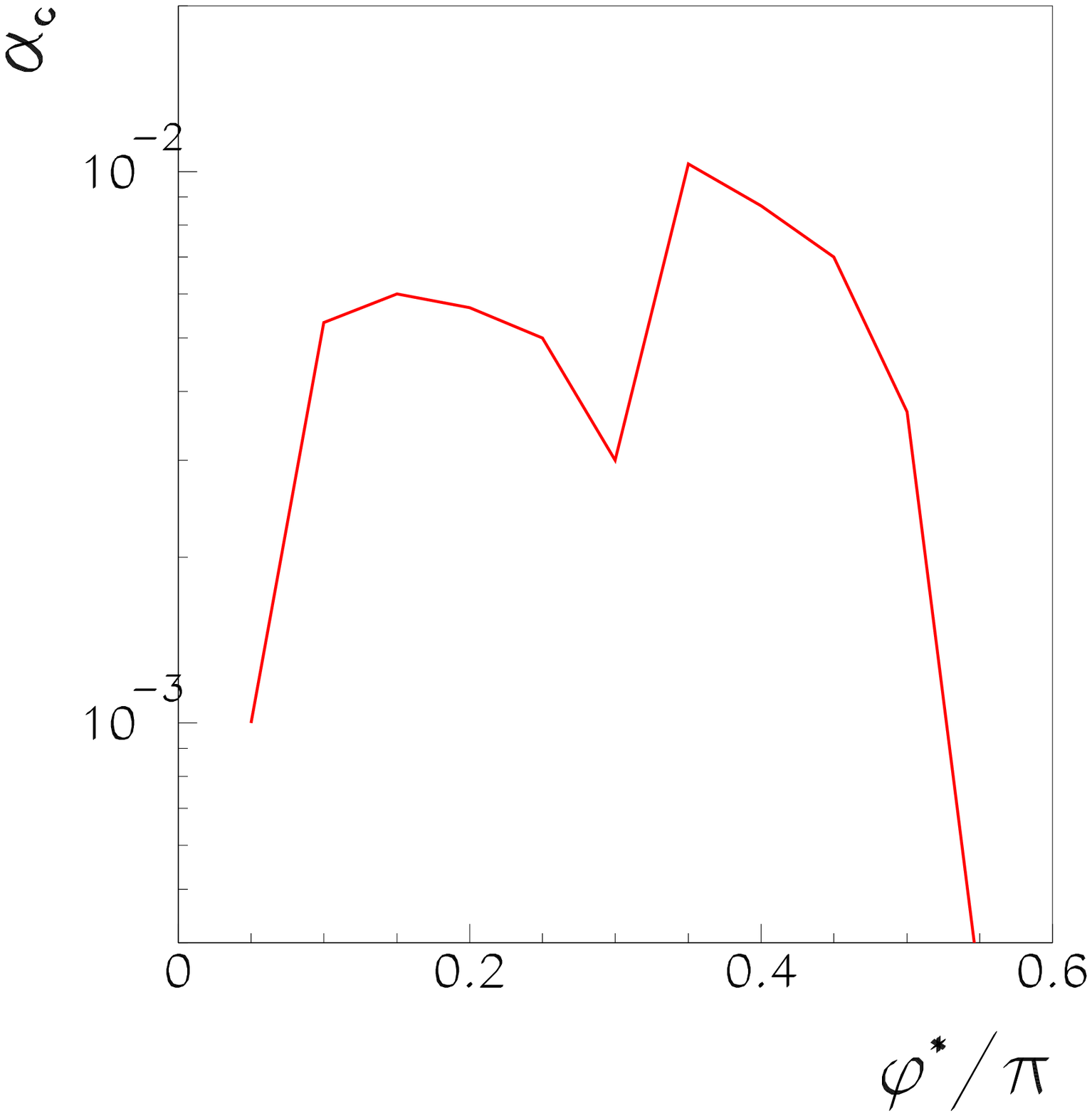}
b) 
\includegraphics[width=5cm]{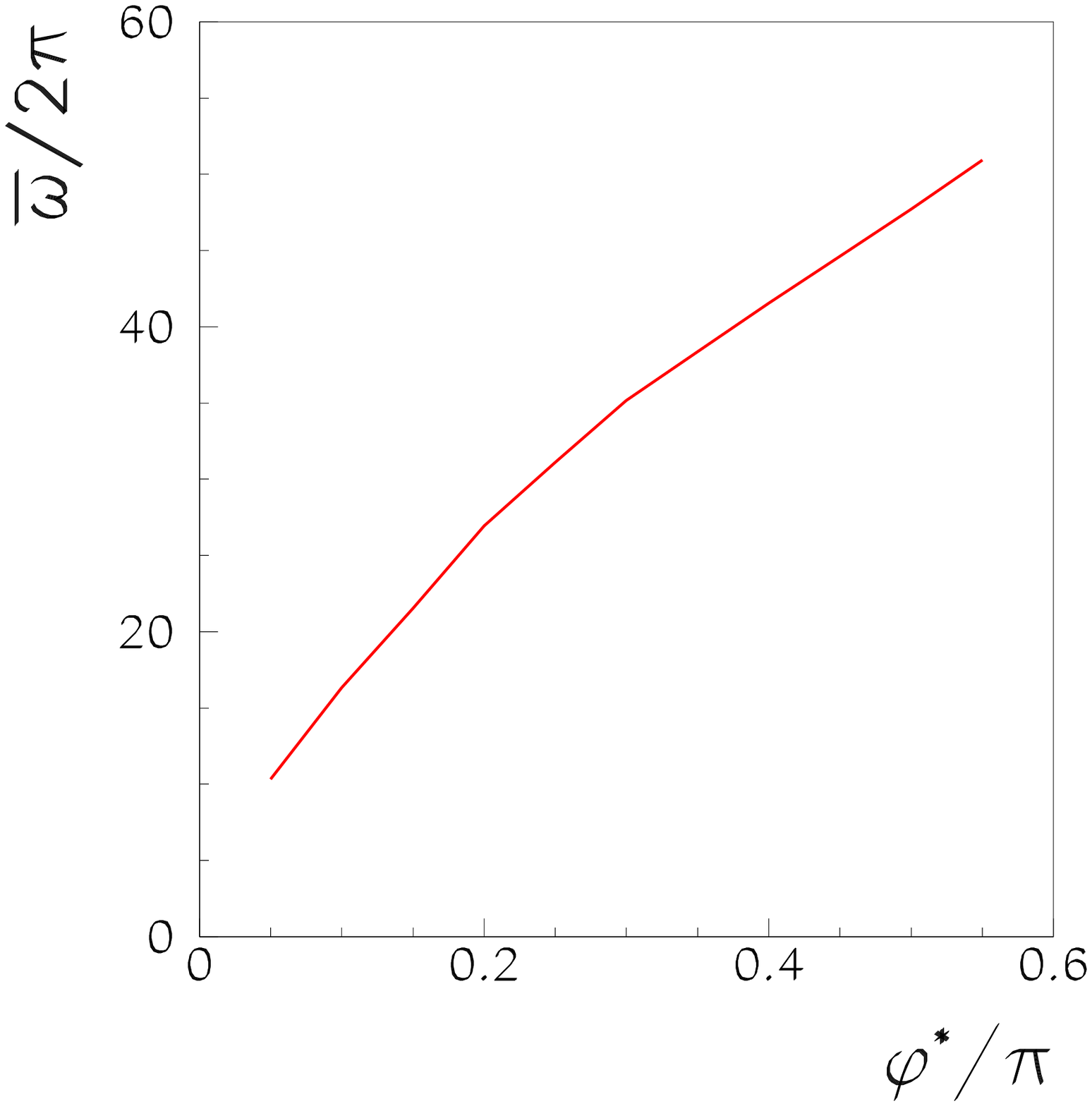}
\end{center}
\caption{%
a) Maximum capacity 
$\alpha_c=P_{\text{max}}/N$ of a network of $N=3000$ fully connected spiking 
IF neurons,
as a function of the learning window asymmetry $\varphi^\ast$.
The spontaneous dynamics induced by a short train of $M=300$  spikes with
$\omega_\mu=20 Hz$ and phases given by encoded pattern $\mu=1$ is
considered, with threshold set to $T=200$. 
b) The output frequency of the replay, as a function of $\varphi^\ast$, when
$P=1$, $T=200$, $N=3000$.
}
\label{fig_spike}
\end{figure}

To estimate the network capacity of the spiking model, we did numerical simulations of the
 IF network in Eqs.\ (\ref{tre}) and (\ref{quattro}),
with $N=3000$ neurons, and connections $J_{ij}$ given by Eq.\ (\ref{eq:learningrule}),
with different number of patterns P.
We give an initial short train of M=300 
spikes chosen at times $t_i= \phi_i^\mu/\omega_\mu$ from pattern $\mu=1$.
To check if the initial train triggers the replay of pattern $\mu=1$ at large times, we measure
the overlap $m^\mu(t)$ between the spontaneous dynamics of the network and the phases of pattern $\mu=1$.
In analogy with Eq.\ (\ref{unobis}), the overlap $m^\mu(t)$ is defined as
\begin{equation}
|m^\mu(t)= \left|\frac{1}{N}\sum_{j=1,\ldots,N} e^{-i 2 \pi t_j^\ast/T^\ast}  e^{i \phi_j^\mu}\right|,
\label{nn}
\end{equation}
where $t_j^\ast$ is the spike timing of neuron $j$ during the spontaneous dynamics, and $T^\ast$ is an estimation of the
 period of the collective spontaneous dynamics.
The overlap in Eq.\ (\ref{nn}) is equal to $1$ when the phase-coded pattern is perfectly retrieved (even though on a different time scale), and is of order $\simeq 1/\sqrt{N}$
when the phases of spikes have nothing to do with the stored phases of pattern $\mu$.
We consider a successful recall each time the overlap (averaged over 50 runs) is larger then $0.7$.
The capacity of the fully connected spiking network as a function of $\varphi^\ast$
is shown in Fig.\ \ref{fig_spike}, for threshold $T=200$ and $N=3000$.
  
While in the analog model the output frequency tends to infinity when
$\varphi^\ast \to \pi/2$, in the spiking model output frequency increases
 but not diverge at
$\pi/2$, as shown in Fig.\ \ref{fig_spike}b.
As in the analog model, the capacity decreases as soon as $\varphi^\ast$ is
near or larger then $\pi/2$. 
With parameters used in Fig.\ \ref{fig_spike}, 
we see that, for $\varphi^\ast$  larger then $0.55 \pi$, 
the stimulation spikes are not able to initiate the recall of the pattern
and the network is silent, unless we choose
a lower threshold. This behavior is similar to a small oscillation
obtained in the analog model at $\varphi^\ast \to \pi/2$.

A systematic study of the dependence of the storage capacity
on the threshold $T$, on the number $M$ of spikes used to
trigger the recall, and on time constants $\tau_m$ and $\tau_s$  still has to be done.
Future work will also consider the case in which patterns to be learned are not
defined through the rate $x_i^\mu$ in Eq. (2), but are defined as sequences of spikes whose timing is exactly given
by $(\phi_i^\mu + 2 \pi n)/\omega_mu$. We expect for this case a higher storage capacity.

\section{Summary and Discussion }
\label{sec_summary}
In this paper we studied the storage and recall of patterns in which
information is encoded in the phase-based timing of firing relative to the cycle.
We analyze the ability of the learning rule given by Eq.\ (\ref{eq:learningrule}) to memorize multiple phase-coded patterns,
such that the spontaneous dynamics of the network, defined by Eq.\ (\ref{uno}),
selectively gives sustained activity which matches one of
the stored phase-coded patterns, depending on the initialization of the network.
It means that if one of the stored items is presented as input to the network at time $t<0$,
and it is switched off at time $t\ge 0$,
the spontaneous activity of the network at $t>0$ gives sustained activity whose phases alignments match those presented before. 
It is not trivial that the network performed retrieval competently, as analog associative memory is hard \cite{treves}.

We compute the storage capacity of phase coded patterns in the analog
model, finding a linear scaling of number of patterns with network size,  with  maximal
capacity $\alpha_c \simeq 0.02$ for the fully connected network.
Our model cannot be easily compared to classical models such as Hopfield's,
since the dynamical phases-coded patterns are strongly different from the
classical static rate-coded patterns.
Whereas the capacity of rate coded networks has been well understood,
that of spiking or phase-coded networks is not yet well understood.
We are not aware of previous work who measures the storage capacity
of  phase-coded patterns, except of the phase-coded patterns in a
spin model by Yoshioka \cite{y}, whose capacity is in quantitative agreement with
our analog model results.
However, the scaling of capacity with connectivity displays properties similar to those of classical models.
For strong dilution ($z\ll N$), the capacity $P_{\text{max}}$, that is the maximum number of patterns encodable in the network,
is proportional to $z$ rather than to $N$. On the other hand, for weak dilution ($z$ of the order of $N$), $P_{\text{max}}$ 
is proportional to the number of neurons $N$.

Our results on the IF spiking model   
show that there is a qualitative agreement between the analog and the spiking model
ability to store multiple phase-coded patterns and recall them selectively.

We also study the storage capacity of the analog model for different degrees
of sparseness and small-worldness of the connections.
We put neurons on the vertices of a two or three dimensional lattice,
and connect each neuron to neurons that are nearer than some distance
in units of lattice spacings.
Then a fraction $\gamma$ of these connections are rewired, deleting the short range connection and creating a 
long range connection to a random neuron.
The existing connections are then defined by the learning
rule Eq.\ (\ref{eq:learningrule}), while other connections are set to zero.

Changing the proportion $\gamma$ between short-range and long-range connections,
we go from a two or three dimensional
network with only nearest-neighbors connections ($\gamma=0$) to a random network ($\gamma=1$).
Small but finite values of $\gamma$ give a ``small world'' topology, similar to that found in many areas
of nervous system.
We see that, for system size $N=24^3$, the capacity of a random network with only $10\%$ connectivity
already has about $40\%$ the capacity of the fully-connected network, showing that there is a saturation effect
when the density of connections grows above $10\%$.

Looking at the dependence on the fraction of short range and long range connections, 
we see that at $z/N=0.11$ the capacity gain with respect to the short range network, given by
$30\%$ long range connections, already is about $80\%$ the gain given by the full random network.
This last factor is likely to increase for larger system sizes,
because the larger the system, the more different are long range connections with respect to short range ones.
This is interesting considering that a long-range connection
clearly have a higher cost then a short-range one, and implies that a small-world network
topology is optimal, as a compromise between the cost of long range connections and the capacity increase.

These results have been found for the analog model in Eq.\ (\ref{uno}).
However, this is not a really spiking model, since only "firing activity" or "probability of spiking" is evaluated.
Therefore we perform numerical simulation of a IF model with spike response kernels, to show that 
while details of dynamics may depend on the model of single unit, the
crucial results of ability to store and recall of phase-coded memories %
can be seen also in a spiking model. Indeed numerical simulations
of the spiking model in Eq. (\ref{tre}) shows competent storage and recall of phase-coded memories.
A pattern of spikes with the stored phases, i.e. with the phase-based timing of spikes relative to the cycle,
is activated by the intrinsic dynamics of the network when  a partial cue of the stored pattern (few spikes at proper timings)
is presented  for a short time.
Fig. \ref{fig_varphi} shows that the network is able to store multiple phase-coded memories and selectively replay one of
them depending from the partial cue that is presented. Changing the parameter $\varphi^\ast$ of
the learning rule changes the period of the cycle during recall dynamics,
while changing the the value of $T$ changes the number of spikes per cycle,
without changing the phase pattern. 
So the same phase-coded pattern can be recalled with one spike per cycle or with a short burst per cycle.  This 
agrees well with recent observations,
like occurrence of phase precession with very low as well as high firing rate \cite{huxter_burgess}.
The number of spikes per cycle is a sort of strength or saliency of recall.
This leaves open the possibility that variations in number of spikes per cycle might convey additional information about other
variables not coded in the phase pattern, as suggested in \cite{huxter_burgess,mate2,yoko_wu} for place cells,
or may convey information about the saliency of
retrieved pattern \cite{mate3}.

In our treatment,  we distinguish a learning mode, in which connections are plastic and  activity is
clamped to the phase-coded pattern to be stored in the synaptic connections $J_{ij}$, from a recall
mode, in which connection strengths do not change. 
Of course, this distinction
is somewhat artificial; real neural dynamics may not be separated so
clearly into such distinct modes. Nevertheless,  data on cholinergic neuromodulatory effects
\cite{Hasselmo,Hasselmo2} in cortical structures
 suggesting that
high levels of acetylcholine
selectively suppress intrinsic but not afferent fiber transmission and enhance long term plasticity,
seems to provide a possible neurophysiological mechanism for this distinction in two operational modes.

Another point is given by the physical constraints on the sign of synaptic connections $J_{ij}$,
that for a given presynaptic unit $j$, are all positive when presynaptic unit $j$ is excitatory,
and all negative when presynaptic unit $j$ is inhibitory,
a condition not respected by our learning formula Eq.\ (\ref{eq:learningrule}) so far.
As a remedy, one may add an initial background weight $J_0$ to each connection, independent of $i$ and $j$,
such that $J_{ij}+J_0>0$ for all $i$ and $j$,
to make all the units excitatory and plastic connections
positive, and then one has to add a global inhibition equal to $-J_0$ times the
mean field $y=\sum_j x_j$ to save equilibrium between excitation and inhibition.
In such a way, we have a network  of $N$ excitatory units, with positive couplings, and a global inhibition,
\begin{equation}
\tau_m \,\dot x_i = -x_i + F \left[\sum_j (J_{ij}+J_0) x_j(t) - J_0 y \right]
\label{due}
\end{equation}
that is mathematically equivalent to Eq.\ (\ref{uno}) when $y=\sum_j x_j$, 
and numerically
also gives the same results when $y=N/2$.

Considering the IF model of Sec.\ \ref{sec_if_model}, also in this case
we can consider a network of all excitatory neurons, with plastic positive connections given by $J_{ij}+J_0$
with $J_0$ constant such that $J_{ij}+J_0>0$,
and then add a proper inhibitory term to save the equilibrium between excitation and inhibition.
In this way, the postsynaptic potential of neuron $i$, after time $t_i$ of its last spike, is given by
\begin{equation}
h_i(t)= \sum_{j/t_j>t_i} (J_{ij}+J_0) \epsilon(t-t_j) - J_0 \sum_{j/t_j>t_i}\epsilon(t-t_j).
\end{equation}
The inhibitory term $- J_0 \sum_{j/t_j>t_i}\epsilon(t-t_j)$
can be realized in different ways, for example imaging that for each excitatory unit $j$, there is
a fast inhibitory interneuron $j$, that emits a spike each time its excitatory unit does, and
is connected with all the other excitatory neurons with constant weight connections $J_0$. 

The task of storing and recalling phase-coded memories has been also
investigated in \cite{14} in the framework of
probabilistic inference.  While we study the effects of couplings
given by Eq.\ (\ref{eq:learningrule}) in the network model Eq.\ (\ref{uno}),
and in a network of IF neurons given by Eq.\ (\ref{tre}), the paper \cite{14} studies
this problem from a normative theory of autoassociative memory,
in which variable $x_i$ of neuron $i$ represents the  neuron $i$ spike
timing with respect to a reference point of an ongoing field
potential,
and the interaction $H(x_i,x_j)$ among units is mediated by the derivative
of the synaptic plasticity rule used to store memories.
In \cite{14}, the case of limited connectivity is studied, showing how
recall performance depends to the degree of connectivity when
connections are cut randomly.
Here we show that performance also depends from the  topology of the connectivity,
and capacity depends not only from the number of
connections but also  from the fraction of long range versus short range connections.

The role of STDP in learning and detecting spatio-temporal patterns has been studied recently in \cite{Masque}.
They show that a repeating spatiotemporal
spike pattern, hidden in equally dense distracter spike trains, can be robustly
detected  by a set of ``listening'' neurons equipped with spike timing-dependent plasticity (STDP).
When a spatio-temporal pattern repeats periodically, it can be considered a periodic phase-coded pattern. 
While in \cite{Masque} the detection of the pattern is investigated when it is the input of the ``listening'' neurons,
in our paper we investigate the associative memory property, which makes the pattern imprinted in the connectivity of
the population an attractor of the dynamics.
When a partial cue of the pattern $\mu$ is presented (or in the analog case the network is initialized with $x_i^\mu(0)$),
then the original stored pattern is replayed. 
Differently from \cite{Masque},  here the pattern $\mu$ is imprinted  in the neural population,
in such a way that exactly the same encoded phase-coded pattern is replayed during persistent spontaneous activity. 
This associative memory behavior, that replay the stored sequence,
 can be a method for recognize an item, by activating the same memorized pattern in response of a similar input,
or may be also a way to transfer the memorized item to another area of the brain (such as for memory consolidation during sleep).

Our results shows that in the spiking model a critical role is played
by the threshold T: changing the
threshold one goes from a silent state to a spontaneously active
phase-coded pattern with one spike per cycle, and then to the same phase-coded
pattern with many spikes per cycle (bursting).
Therefore, a possible conjecture is that when we observe the replay of a
periodic spatiotemporal pattern,
then if one is able to change the threshold of spiking of the neurons,
one can observe the same phase pattern but with more or less spikes
per cycle.

In future we will also investigate the capacity when patterns with
different frequencies are encoded in
the same network. In this paper we compute the capacity when encoded
patterns have all the same frequency,
however it would be interesting to see how the network work with
different frequencies.
Preliminary results on the analog model \cite{Erice}
show that it's possible to store two different frequencies in a
manner that both are stable if
a relationship holds between the two frequencies and the shape of the
learning window.
We plan to investigate this both in the analog and in the spiking model.

Concerning the shape of learning kernel $A(\tau)$, in our model the values of the connections $J_{ij}$ depend on the kernel shape only through the time
integral of the kernel, and  the Fourier transform of the kernel at the frequency of the encoded pattern.
As shown in our previous paper \cite{PREYoshioka}, best results are obtained when
the time integral of the kernel is equal to zero $\int_{-\infty}^\infty  A(\tau) d\tau= \tilde A(0)=0$.
This choice will assure the global balance between excitation and inhibition.
Hence in the present paper we choose to set this value to zero, and varied only the phase $\varphi^\ast$ of the Fourier transform. 
 The case of a purely symmetric kernel corresponds to $\varphi^\ast=0$, 
while a purely
anti-symmetric (causal) one corresponds to $\varphi^\ast=\pi/2$.
We find that intermediate values $0<\varphi^\ast<\pi/2$ work better then
values $\pi/2<\varphi^\ast<\pi$, for which the network is not able to replay
patterns competently. 
We study here positive values of $\varphi^\ast$, i.e. a causal kernel
(potentiation for pre-post, and depotentiation for post-pre), however 
an anti-causal kernel  would give a
negative value of $\varphi^\ast$, and a negative frequency of replay, that 
is the pattern
would be replayed in time reversed order \cite{PREYoshioka,Erice}. 

For what concerns the variety of shapes of the learning window observed in the
brain, it has to be considered that they may accomplish different needs, 
such as maximize storage capacity, or set the time scale of replay, which in our model depends on the kernel shape.
Notably, in our model we observe a good 
storage capacity for values of phase $\varphi^\ast$ 
which correspond to a large interval of frequency of oscillation during replay, and
therefore there is good storage capacity also for 
 values of $\varphi^\ast$ which
 correspond to a compressed-in-time replay of phase-pattern 
on very short time-scale.
Future work will investigate the relation between this framework and
the time-compressed replay of spatial experience in rat. 
It has been indeed observed that during pauses in exploration and during
sleep, ensembles of place cells in the rat
hippocampus and cortex re-express firing sequences corresponding to past spatial
experience \cite{replay1,replay2,replay3,replay4,replay5}. 
Such time-compressed hippocampal replay of behavioral sequences 
co-occurs with ripple events:
 high-frequency %
oscillations that are associated with increased hippocampal-cortical
communication. It's intriguing the hypothesis that 
sequence replay during ripple  is the recall of one of the multiple
stored phase-coded
memories, a replay that occur on a fast time scale (high $\bar\omega$), 
triggered 
by a short sequence of spike.

\end{document}